\documentclass[aps,10pt,pra,twocolumn,amssymb,amsmath,amsfonts,showpacs,
floatfix,citeautoscript,preprintnumbers,footinbib,superscriptaddress
]{revtex4-1}
\usepackage{times}
\usepackage{graphicx}
\usepackage[latin1]{inputenc}

\newcommand{\moy}[1]{\langle{#1}\rangle}

\newcommand{\up}{\uparrow}
\newcommand{\down}{\downarrow}
\newcommand{\dx}{\partial_x}

\begin{document}

\title{Multimer formation in 1D two-component gases and trimer phase in the asymmetric attractive Hubbard model}
\author{Guillaume Roux}\email{guillaume.roux@u-psud.fr}
\affiliation{Laboratoire de Physique Th\'eorique et Mod\`eles statistiques, Universit\'e Paris-Sud, CNRS, UMR8626, 91405 Orsay, France.} 
\author{Evgeni Burovski}
\affiliation{Physics Department, Lancaster University, Lancaster LA1 4YB, UK.}
\author{Thierry Jolic{\oe}ur} 
\affiliation{Laboratoire de Physique Th\'eorique et Mod\`eles statistiques, Universit\'e Paris-Sud, CNRS, UMR8626, 91405 Orsay, France.}

\date{\today}

\begin{abstract}
  We consider two-component one-dimensional quantum gases at special
  imbalanced commensurabilities which lead to the formation of
  multimer (multi-particle bound-states) as the dominant order parameter. 
  Luttinger liquid theory supports a mode-locking mechanism in which 
  mass (or velocity) asymmetry is identified
  as the key ingredient to stabilize such states. While the scenario
  is valid both in the continuum and on a lattice, the effects of
  umklapp terms relevant for densities commensurate with the
  lattice spacing are also mentioned. 
  These ideas are illustrated and confronted with the physics
  of the asymmetric (mass-imbalanced) fermionic Hubbard model with
  attractive interactions and densities such that a trimer phase can
  be stabilized. Phase diagrams are computed using density-matrix
  renormalization group techniques, showing the important
  role of the total density in achieving the novel phase. The
  effective physics of the trimer gas is as well studied. Lastly, the
  effect of a parabolic confinement and the emergence of a crystal
  phase of trimers are briefly addressed. This model has connections
  with the physics of imbalanced two-component fermionic gases and
  Bose-Fermi mixtures as the latter gives a good phenomenological 
  description of the numerics in the strong-coupling regime.
\end{abstract}

\maketitle

\section{Introduction}

The notions of quantum liquids and their instabilities are
paradigmatic for condensed matter physics~\cite{Leggett2006}. For
multicomponent fluids, an important set of instabilities is associated
with interactions between components. A classic example is the Cooper
instability of a spin-$1/2$ Fermi liquid~: even an infinitesimal
attractive coupling between fermions of opposite spins drives a phase
transition into the Bardeen-Cooper-Schrieffer
superconductor~\cite{Schrieffer1999}. A one-dimensional (1D)
counterpart of the Fermi liquid, the spinful Luttinger liquid, has a
similar instability, where an attractive inter-spin coupling opens a
gap in the spin channel~\cite{Giamarchi1995, Giamarchi2004}.

Traditionally, the bulk of the discussion on two-species liquids
assumed the SU(2) spin symmetry. The recent years have witnessed a growing availability of experimental
studies of mixtures of unlike particles. This includes loading
ultracold atoms to spin-dependent optical lattices~\cite{Mandel2003}, and
trapping atoms of different masses~\cite{Wille2008}
or even different statistics~\cite{Giorgini2008}. While most of
experimental progress so far is in the domain of ultracold atoms, we
stress that the relevance of such \emph{asymmetric} mixtures is not
confined to the realm of cold gases~: dealing with more traditional
solid state systems, one faces an asymmetric mixture situation as soon
as the Fermi level spans several bands (which \textit{a priori} need
not be equivalent). This setup is typical for such diverse materials
as semi-metallic compounds, mixed-valence materials, organic
superconductors~\cite{Penc1990}, small radius nanotubes
\cite{Carpentier2006}, and even graphene-based
heterostructures~\cite{Martin2008}.

A generic question immediately arises~: given a two-component mixture,
what is the role of (the lack of) SU(2) symmetry? Or, more precisely,
does the symmetry between components limit the set of instabilities of
a liquid? Clearly, the answer might depend on the universality
class of the liquid, and on the particular way the symmetry is
broken. The simplest albeit non-trivial way of breaking the symmetry
is to assume species-dependent masses of the particles. Even if we
consider the few-body problem, this is known to bring new physics like
the Efimov phenomenon~: while for an equal-mass Fermi liquid the only
allowed bound state is a Cooper pair, three-body bound states
(trimers) appear once the mass ratio exceeds a certain
threshold~\cite{Braaten2006}. 
The atom-dimer scattering is strongly affected by
the mass asymmetry ~\cite{Petrov2003}
and the ultimate fate of a
Fermi liquid in presence of the Efimov effect is currently an open
question being actively investigated~\cite{Levinsen2009}. 
All these theoretical considerations are strongly motivated by cold-atom experiments which have recently achieved
degeneracy of Fermi gases with different masses~\cite{Wille2008}
and spin-imbalanced two-component fermionic gases~\cite{Partridge2006}.

\begin{figure}[b]
\centering
\includegraphics[width=\columnwidth,clip]{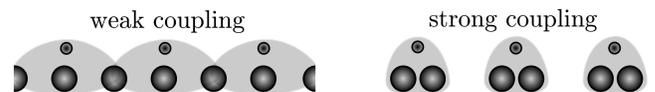}
\caption{Qualitative formation of trimers in the weak-coupling picture
  of the bosonization (on the left) and in the strong-coupling picture for
  large interactions (on the right).}
\label{fig:trimers}
\end{figure}

The physics of 1D quantum many-body systems offers powerful methods~\cite{Giamarchi2004},
both analytical and numerical, to have quantitative predictions on the fate of the
 Luttinger liquid in the presence of perturbations.
The role of mass asymmetry for a two-component Luttinger liquid has
been investigated in the renormalization group (RG) framework
originally in the context of solid state physics~\cite{Muttalib1986, Penc1990, Loss1994}, 
and recently revisited mostly in the context of cold atoms~\cite{Cazalilla2003, Mathey2004, Cazalilla2005, Mathey2007,
  Mathey2007a, Lu2009, Crepin2010, Orignac2010, Tsvelik2010}, and
supplemented by numerical investigations~\cite{Pollet2006, Mering2008, Batrouni2009, Wang2009}. 
Overall the consensus was that the only new instability arising
due to asymmetry is the collapse (demixing) instability for large
asymmetry and/or strong interspecies attraction (repulsion). Recently,
a novel family of instabilities was predicted~\cite{Burovski2009} to
exist due to the interplay between \emph{polarization and asymmetry}~:
These instabilities only take place for \emph{polarized} mixtures of
either statistics, and are characterized by the locking of the ratio
of the densities to a \textit{rational} value. Subsequent work in
Ref.~\onlinecite{Orso2010} elucidated the relation of these
instabilities and existence of few-body bound states. A qualitative
picture of the mode-locking mechanism and the strong-coupling limit of
the trimer formation is given in Fig.~\ref{fig:trimers}. The latter
regime recalls another approach to multi-particle bound-states which
is the use of many-colors ($N$-component) fermions~\cite{Wu2005, Capponi2008, Roux2009} 
with which the physics of the trimers share 
qualitative features.

This paper is divided in two main parts : the first one investigates
in detail the bosonization approach and the mode-locking mechanism
mentioned above, while the second is dedicated to the specific but
important example of the 1D asymmetric Hubbard model
using the density-matrix renormalization group (DMRG) 
technique~\cite{White1992}. The
predictions of the first part account for most of the numerical data, but a
more phenomenological Bose-Fermi picture is proposed as a
complementary analysis. 
Other important questions such as the effect
of a trapping potential or the emergence of crystal phases are
eventually addressed.

\section{Bosonization analysis}
\label{sec:boso}

In this section, we describe the salient features of the effective
bosonic field theory appropriate to a 1D mixture of two distinct
fermionic (or bosonic) atoms. The aim of this section is to give a
bosonization interpretation for the formation of few-body bound states
and their effective behavior through a mode-locking mechanism between
the two species. Predictions on the nature of the resulting phase are
then made. The theory is a priori valid for models in the continuum or
the continuous version of lattice models at generic
(i.e. non-commensurate) densities. The effects of the presence of the
lattice on certain commensurate densities will be briefly discussed in
section~\ref{sec:lattice}. Notation conventions are standard and taken from 
Ref.~\onlinecite{Giamarchi2004}.

\subsection{Mode locking mechanism}
\label{sec:modecoupling}

The two species are labeled by a pseudo-spin index $\sigma =
{\up,\down}$ and their corresponding densities $n_{\sigma}$ such that
$n=n_{\up}+n_\down$ is the total density. Each species can be
described by a scalar field $\phi_\sigma$ and its dual 
$\theta_\sigma$. The creation operators can be expressed
as a function of these fields, with, for fermions
\begin{equation}
\label{eq:fermions}
\Psi^\dagger_\sigma(x) \sim \Big( n_{\sigma}- \frac{1}{\pi} \dx\phi_\sigma
\Big)^{1/2} \sum_{p} e^{ i (2p+1)(k_{\sigma} x -\phi_\sigma)} \, e^{-i\theta_\sigma}\,,
\end{equation}
and, for bosons,
\begin{equation}
\label{eq:bosons}
b^\dagger_\sigma(x) \sim \Big( n_{\sigma}- \frac{1}{\pi} \dx\phi_\sigma
\Big)^{1/2} \sum_{p} e^{ i 2p (k_{\sigma} x-\phi_\sigma) } \, e^{-i\theta_\sigma}\;.
\end{equation}
We have included all higher harmonics : as a consequence, the summation
is over all integers~\cite{Haldane1981}. The ``Fermi momenta'',
$k_\sigma=\pi n_\sigma$ are a priori not equal to each other,
corresponding to a spin-imbalanced situation. The
density operators $\hat{n}_\sigma$ read
\begin{equation}
\label{HaldaneN}
\hat{n}_\sigma(x) \sim \Big( n_{\sigma}- \frac{1}{\pi} \dx\phi_\sigma\Big)
\sum_p e^{ i2p(k_\sigma x -\phi_\sigma)} \; .
\end{equation}
The effective low-energy Hamiltonian can be written in terms of the
fields $\phi_\sigma$ and their canonically conjugate momentum
$\Pi_\sigma = \dx \theta_\sigma/\pi$. In the case of absence of
inter-species interactions the effective bosonic theory is given by
$\mathcal{H}_0(\phi_\up) +\mathcal{H}_0(\phi_\down)$, where
\begin{equation}
\mathcal{H}_0(\phi_\sigma) = \dfrac{v_\sigma}{2\pi} \int\! dx\,\left[
K_\sigma(\pi\Pi_\sigma)^2 +K_\sigma^{-1} \left(\dx\phi_\sigma\right)^2 \right]\; ,
\label{H0}
\end{equation}
where $v_\sigma$ is the sound velocity and $K_\sigma$ the so-called
Luttinger parameter, which is equal to one in the free fermions or
free hard-core bosons cases. Taking into account density-density
interactions between species, of the generic form $\int\! dx dx'\,
U(x-x') \hat{n}_\up(x) \hat{n}_{\down}(x')$, changes the effective
theory and brings new kinds of terms : zero-momentum terms in the
density representation \eqref{HaldaneN} couples the two spin species
through a bilinear operator
\begin{equation}
\label{H1}
\mathcal{H}_1 = \frac{g}{2\pi} \int\! dx \, (\dx \phi_\up)(\dx \phi_\down) \;,
\end{equation}
where $g$ is a forward scattering constant, and higher harmonics terms
involving multiples of the spatial frequencies $k_\sigma$ :
\begin{align}
\label{H2}
\sum_{p,q>0}G_{pq}^{\pm}\!\int\!dx\,\cos{\big[2\pi(pn_\up \pm qn_\down )x - 2(p\phi_\up \pm q\phi_\down)\big]}
\end{align}
where $G_{pq}^{\pm}$ are non-universal coupling constants. Clearly,
if the generalized commensurability condition
\begin{equation}
\label{pq}
p\,n_\up - q n_\down =0 \;,
\end{equation}
is satisfied (with $p$ and $q$ coprime integers), and provided these
terms are relevant, they will tend to lock the up and down fields
together. When the densities are fine-tuned to the definite
commensurability \eqref{pq}, then all other cosine operators in the
sum are oscillating, in which case they don't contribute in the
continuum limit (or they are less relevant for multiples of $p$ and
$q$). The remaining important operator in the sum \eqref{H2} is thus
the sine-Gordon term
\begin{equation}
\label{H2pq}
\mathcal{H}_2 = G \int\! dx \, \cos{\sqrt{8}\phi_a} \; ,
\end{equation}
with the combination
\begin{equation}
\label{eq:phia}
\phi_a=\frac{1}{\sqrt{2}}(p\phi_{\up}-q\phi_{\down})\;.
\end{equation}
For attractive interactions $G<0$ (we will argue below that this
choice favors the relevance of the term), energy will be minimized
when the field is pinned to $\moy{\phi_a} = 0$. Notice again that the above
argument on the mode-locking mechanism does not rely on the presence
of a lattice. Lastly, the cosine locks a combination of the bosonic
modes but at a generic total density $n$, there remains another bosonic
mode leaving the full excitation spectrum gapless. We will see that the
latter describes the effective behavior of the bound-states. In the
following, we dub $\phi_b$ this massless bosonic mode.

We can draw a last remark on the operators \eqref{H2} : they
have high scaling dimensions near the free fermion fixed point
and are expected to be irrelevant apart from some special circumstances
which are the object of this work. In the fermionic language, they involve
$(p+q)$-body interactions of the form:
\begin{equation}
\sum_{\{k\}} \prod_{i=1}^q  \psi^\dagger_{R\down}(k_i) \psi_{L\down}(k_i') \, 
\prod_{j=1}^{p} \psi^\dagger_{L\up}(k_j'') \psi_{R\up}(k_j''') + \mathrm{h.c.} \; ,
\label{H2pqFERM}
\end{equation}
where the summation over $\{k\}$ runs over all combinations of
$2(p+q)-1$ momenta due to the total momentum conservation law~:
$\sum_{i=1}^q ( k_i - k_i') + \sum_{j=1}^p (k_j'' - k_j''') =0$. 
Such interactions appear at high order in perturbation theory in a Hubbard model
for example or after several steps of a RG treatment.
For practical purposes it is simpler to work with the bosonic formulation
given by Eq.~\eqref{H2pq}, and this is what we do from now on.

In the following, we assume that the densities are commensurate via the condition
\eqref{pq}, and analyze the simplified effective theory written in terms of the $\up$ and $\down$ fields
\begin{equation}
\mathcal{H}= \mathcal{H}_0(\phi_\up) + \mathcal{H}_0(\phi_\down) + \mathcal{H}_1 + \mathcal{H}_2 \; ,
\label{fullmodel}
\end{equation}
where the velocities $v_\sigma$ and Luttinger parameters $K_\sigma$
are determined by the intra-species interactions.  
The quadratic part $\mathcal{H}_0(\phi_\up) + \mathcal{H}_0(\phi_\down) + \mathcal{H}_1$
can be diagonalized by a Bogoliubov transformation~\cite{Engelsberg1964}
which could give a starting point for a perturbative RG calculation~\cite{Penc1990,
Cazalilla2003, Cazalilla2005, Mathey2007, Crepin2010}. Due to the
velocity asymmetry, additional couplings are generated and velocities
are renormalized along the flow. The discussion of the nature of the 
gapped phases and their correlations remains unclear.
In particular, diagonalizing the
quadratic part $\mathcal{H}_0(\phi_\up) + \mathcal{H}_0(\phi_\down) +
\mathcal{H}_1$ of the Hamiltonian \eqref{fullmodel} does not give, 
apart from special choice of the parameters, 
the combination \eqref{eq:phia} that appears in the
cosine term $\mathcal{H}_2$. In the next section, we take the following
strategy : we look for the conditions under which the
quadratic part and the cosine term are simultaneously
diagonalizable. At the price of a restriction on the parameters range, 
the analysis can be done safely both for the criteria of 
relevance of the cosine and for the correlation functions in the single-mode
phase. In spite of the limitation of the approach, we believe the scenario
does occur without this restriction : 
as shown numerically in Sec.~\ref{sec:trimers} on a realistic model,
the single-mode multimer phase can span a wide region of the phase diagram.

Lastly, we notice that, similarly to the phase separation
criteria in two-component mixtures (when one of the mode velocities vanishes), 
the single-mode phase will undergo a phase separation instability
when the gapless mode velocity $v_b$ vanishes. We thus expect to find
the single-mode phase surrounded with the two-mode phase and a 
demixed phase.

\subsection{Field transformation}
\label{sec:fieldtransfo}

We have qualitatively discussed the fact that the physics should
generically be described by two fields
$\phi_s$ where $s=a,b$, with $\phi_a =
(p\phi_{\up}-q\phi_{\down}) / \sqrt{2}$ being the one entering in the
cosine term \eqref{H2pq}. In general, it is hard to have a complete
form for the transformation between the $\phi_s$ and the
$\phi_{\sigma}$. Such a transformation is important
both for the RG analysis and the calculation of physical correlators
which are naturally expressed in terms of the $\phi_{\sigma},
\theta_{\sigma}$ fields. Below, we discuss a special case where the
transformation can be performed and its range of validity.

The simplest transformation, and yet rather general, one can work with
is a linear combination of the fields with coefficients that are
independent of the position~:
\begin{align}
 \phi_{\up}   &= \mathfrak{p}_{a\up} \phi_a + \mathfrak{p}_{b\up} \phi_b\,,     
& \theta_{\up}   &= \mathfrak{t}_{a\up} \theta_a + \mathfrak{t}_{b\up} \theta_b\,, \\
 \phi_{\down} &= \mathfrak{p}_{a\down} \phi_a + \mathfrak{p}_{b\down} \phi_b\,, 
& \theta_{\down} &= \mathfrak{t}_{a\down} \theta_a + \mathfrak{t}_{b\down} \theta_b\,.
\end{align}
When $p \neq q$, excitations corresponding to the eigenmodes
$\phi_{a,b}$ carry both spin and charge modes which are respectively
the sum and the difference of the $\up$ and $\down$ modes. As the
transformation must preserve the commutation relations
\begin{align}
[ \theta_{\sigma}(x), \nabla   \phi_{\sigma'}(x') ] &= i\pi\delta_{\sigma\sigma'} \delta(x-x') \,,\\
[   \phi_{\sigma}(x), \nabla \theta_{\sigma'}(x') ] &= i\pi\delta_{\sigma\sigma'} \delta(x-x') \,,
\end{align}
we get that $\mathfrak{p}_{a\sigma} \mathfrak{t}_{a\sigma'} +
\mathfrak{p}_{b\sigma} \mathfrak{t}_{b\sigma'} =
\delta_{\sigma\sigma'}$. Then, we have
\begin{align}
 \mathfrak{t}_{a\up} &= \mathfrak{p}_{b\down}/D\,, & \mathfrak{t}_{a\down} &= -\mathfrak{p}_{b\up}/D\,,\\
 \mathfrak{t}_{b\up} &=-\mathfrak{p}_{a\down}/D\,, & \mathfrak{t}_{b\down} &= \mathfrak{p}_{a\up}/D\,,
\end{align}
with the determinant 
\begin{equation}
D = \mathfrak{p}_{a\up} \mathfrak{p}_{b\down} -
\mathfrak{p}_{a\down} \mathfrak{p}_{b\up} =
 (\mathfrak{t}_{a\up} \mathfrak{t}_{b\down} -\mathfrak{t}_{a\down} \mathfrak{t}_{b\up})^{-1}\,.
\end{equation}
In a shortened version, we have $\phi_{\sigma} = \mathcal{P} \phi_s$,
where $\mathcal{P}$ is the matrix of the $\mathfrak{p}$, and
$\theta_{\sigma} = (\mathcal{P}^{-1})^{t} \theta_s$. If $\mathcal{P}$
is unitary, the $\theta$ and the $\phi$ undergo the same
transformation.

We now impose that $\phi_a = (p\phi_{\up}-q\phi_{\down}) / \sqrt{2}$
which gives
\begin{align}
\label{eq:tas}
\mathfrak{t}_{a\up} &= \frac{p}{\sqrt{2}}\;, & \mathfrak{t}_{a\down} &= -\frac{q}{\sqrt{2}}\,.
\end{align}
As we want to cancel the cross-terms in Eq.~\eqref{fullmodel}, we
require that~:
\begin{align}
  v_{\up} K_{\up} \mathfrak{t}_{a\up} \mathfrak{t}_{b\up} &= -v_{\down}
 K_{\down} \mathfrak{t}_{a\down} \mathfrak{t}_{b\down} \,,\\
  \frac{v_{\up}}{K_{\up}} \mathfrak{p}_{a\up} \mathfrak{p}_{b\up} + 
\frac{v_{\down}}{K_{\down}} \mathfrak{p}_{a\down}
  \mathfrak{p}_{b\down} &= -g(\mathfrak{p}_{a\down} \mathfrak{p}_{b\up} 
+ \mathfrak{p}_{a\up} \mathfrak{p}_{b\down})\,,
\end{align}
which can be rewritten as~:
\begin{eqnarray}
p v_{\up} K_{\up} \mathfrak{t}_{b\up}\quad -& q v_{\down} K_{\down} &\mathfrak{t}_{b\down} =0 \\
-\Big(p\frac{v_{\down}}{K_{\down}}+gq\Big)\mathfrak{t}_{b\up}\quad + &\displaystyle 
\Big(q \frac{v_\up}{K_\up}+gp\Big)&\mathfrak{t}_{b\down} = 0\;.
\end{eqnarray}
There exists a non-zero solution only if the condition~:
\begin{equation}
\label{eq:condition}
v_{\up}\Big(v_{\up}+\frac{gp}{q}K_\up\Big) = v_{\down}\Big(v_{\down}+\frac{gq}{p}K_\down\Big) 
\end{equation}
is satisfied. When this condition is satisfied, we have a
one-parameter family of transformations with the desirable property of
having only one eigenmode in the argument of the cosine operator.  The
parameter is just the choice of scale of the field $\phi_b$~: in 1D, we
can change the scale of the Bose field provided we change accordingly
its Luttinger parameter $K_b$. Here, we choose the scale of $\phi_b$ so
that~:
\begin{align}
\mathfrak{t}_{b\up}   &= \frac{q}{\sqrt{2}}\sqrt{\frac{v_{\down}}{v_\up}} K_{\down}\;, &
\mathfrak{t}_{b\down} &= \frac{p}{\sqrt{2}}\sqrt{\frac{v_{\up}}{v_\down}}K_{\up}\,,
\end{align}

The condition \eqref{eq:condition} strongly reduces the range of
applicability of the transformation : for a given coupling $g$, the
Luttinger parameters and velocities of each species must satisfy the
above relation.
When the transformation can be used, \eqref{fullmodel} splits into a
free boson field for $b$ and a sine-Gordon model for $a$ :
$\mathcal{H} = \mathcal{H}_0(\phi_b) +
\mathcal{H}_{\text{sG}}(\phi_a)$ with $\mathcal{H}_{\text{sG}} =
\mathcal{H}_0 + \mathcal{H}_2$. In this case, the new velocities and
Luttinger parameters associated with the $a,b$ modes are given by the
following relations~:
\begin{align}
\nonumber
v_a^2 &= \frac{\sqrt{v_{\up}v_{\down}}}{2\mathcal{K}} \Big(p^2 v_\up 
K_{\up}\frac{v_\up}{v_\down} + q^2 v_\down K_{\down} 
\frac{v_\down}{v_\up} - gpq K_{\up}K_{\down}\Big) \,,\\
\label{eq:Ka}
K_a &= \mathcal{K}\frac{\sqrt{v_{\up}v_{\down}}}{v_a} \,, \\
\nonumber
v_b^2 &= \frac{\sqrt{v_{\up}v_{\down}}}{2\mathcal{K}}\left(p^2 v_{\down} 
K_{\up} + q^2 v_{\up} K_{\down} + gpq K_{\up}K_{\down}\right)\,, \\
\nonumber
K_b &= K_{\up}K_{\down}\mathcal{K}\frac{\sqrt{v_{\up}v_{\down}}}{v_b}\,.
\end{align}
where we have defined~:
\begin{equation}
 \mathcal{K} = \frac{p^2 v_{\up} K_{\up} + q^2 v_{\down} K_{\down}}{2\sqrt{v_{\up}v_{\down}}}\,.
\end{equation}
With our definition of $\phi_a$ and provided the sine-Gordon
description is applicable, the requirement for the cosine to be
relevant, and thus to enter the single-mode phase, is simply 
\begin{equation}
K_a < 1\;.
\end{equation}
One qualitatively observes that a velocity much smaller than the
other favors a small $K_a$ and that
large attractive interactions $g<0$ will help increase $v_a$ and
reduce $K_a$. In the following, we consider limiting cases in which
the discussion simplifies in order to identify how the parameters
would favor the formation of a gap in the $a$ sector.

The limit $v_b^2=0$ (attained with attractive interactions) signals the 
transition to the phase-separated or Falicov-Kimball regime from the 
multimer phase.

\subsubsection{The case of equal velocities}

When $v_\up=v_\down=v_0$, the condition \eqref{eq:condition} imposes
that either (i) $g=0$ or (ii) $\displaystyle \frac{K_\up}{K_\down} =
\frac{q^2}{p^2}$. The transformation and new velocities and Luttinger
parameters then take a simple form : in case (i), we have
$v_a=v_b=v_0$ and~:
\begin{align}
\nonumber
\phi_a &=\frac{1}{\sqrt{2}}(p\phi_\up - q\phi_\down)\,, &\phi_b &=\frac{1}{\sqrt{2}}\left(qK_\down 
\phi_\up + pK_\up\phi_\down\right)\,, \\
\label{eq:eqv(i)}
K_a &= \frac{p^2 K_\up +q^2 K_\down}{2}\,, & K_b &= K_\up K_\down \frac{p^2 K_\up +q^2 K_\down}{2}\,.
\end{align}
while in case (ii), we have $\mathcal{K} = p^2K_\up$ and
\begin{align}
\nonumber
\phi_a &=\frac{1}{\sqrt{2}}(p\phi_\up - q\phi_\down)\,, & \phi_b &=\frac{pK_\up}{q\sqrt{2}}
\left(p \phi_\up + q\phi_\down\right)\,, \\
\nonumber
v_a^2 &=v_0^2\Big(1-\frac{gp}{2qv_0}K_\up\Big)\,, & v_b^2 &= v_0^2\Big(1+\frac{gp}{2qv_0}K_\up\Big)\,,\\
\nonumber
K_a &= p^2 K_\up & K_b &= \frac{p^4K_\up^3 }{q^2\sqrt{1+\frac{gp}{2qv_0}K_\up}}\,.
\end{align}
In both cases, having $K_a<1$ would require a very small $K_{\up}$
(assuming $q=1$ for example). This could be realized with long-range
intra-species interactions but may not be easily achievable. Notice a
peculiarity of the formula for the massive mode $\phi_b$ in
\eqref{eq:eqv(i)}: while $K_\up$ and $K_\down$ are length-scale
dependent (in the RG sense), the expression in \eqref{eq:eqv(i)} holds
on all length-scales.

\subsubsection{The limit of large asymmetry}

In order to identify the influence of the velocities ratio on $K_a$, one can introduce the dimensionless
quantities $\nu = v_\down/v_\up$, $\rho = q/p$ and $\gamma =
g/v_\up$. Then, \eqref{eq:condition} and \eqref{eq:Ka} are rewritten
as
\begin{align}
\label{eq:condition-simple}
1+\rho \gamma K_\up &= \nu(\nu+\gamma \rho^{-1} K_\down) \,, \\
\label{eq:Ka-simple}
K_a &= \nu \frac{K_\up + \rho\nu K_\down}{K_\up - \rho\gamma K_\up K_\down \nu + \rho^2K_\down \nu^3 }\,.
\end{align}
If one takes into account \eqref{eq:Ka-simple} only, $K_a$ vanishes in
the limit of large velocity ratio $\nu \rightarrow 0$ or $\infty$
and passes through a maximum in between so that there are two windows
of $\nu$ such that $K_a<1$. The smaller the maximum, the wider these
windows are so, clearly, negative and large interactions ($\gamma <
0$) favor the mode-locking mechanism. Yet, \eqref{eq:condition-simple}
imposes another constraint and we just consider the $\nu \rightarrow
0$ limit for simplicity. There, this limit is possible provided $K_\up
\simeq -1/\gamma \rho$, i.e. in the case of attractive interaction
only. As a consequence, this analysis shows that we should expect the
formation of multimer in the attractive and large interaction regime,
favored by large asymmetry.

\subsubsection{In the single-mode Luttinger liquid phase}

Deep in the massive-$a$ phase, one can make a crude quadratic
approximation to the cosine operator in \eqref{H2pq} by replacing it
with a mass term $\propto (p\phi_\up-q\phi_\down)^2$. This leads to
approximate expressions for the velocity and Luttinger parameter of
the remaining mode~$b$\ :
\begin{align*}
 v_b^2 & = v_\up v_\down \frac{p^2 K_\up v_\down + q^2 K_\down v_\up}{p^2
K_\up v_\up + q^2 K_\down v_\down}\; , \\
K_b & = \frac{1}{2}K_\up K_\down \frac{(p^2 K_\up + q^2 K_\down)^2} {\sqrt{p^2
\frac{K_\up}{v_\up}+q^2 \frac{K_\down}{v_\down}} \sqrt{p^2
{K_\up}{v_\down}+q^2 {K_\down}{v_\up}}} \;,
\end{align*}
which reduces to the correct result for equal velocities.

\subsection{Correlation functions and the nature of the phases}
\label{subsec:corr}

In one-dimensional models, the classification of the groundstates is
determined by their dominant correlations. One can break discrete
symmetries (for instance translational symmetry on a lattice model)
but order parameters associated with continuous symmetries are always
zero. The naming of a phase then corresponds to the connected
equal-time correlator with the slowest decay in space. Quite
generally, these correlators are asymptotically decaying either
algebraically or exponentially. Such algebraic correlations are
usually referred to as a quasi long-range order (QLRO). The slowest
decay (or smallest exponent of algebraic correlations) criteria is
based on a RPA argument by considering a set of weakly coupled
Luttinger liquids~\cite{Giamarchi2004} which shows that order will
build up provided the exponent of the correlator is smaller than two,
and that the main instability is associated with the smallest
exponent. However, if the Green's function, associated with
$\Psi_\sigma$ which is not an order parameter, has the slowest
decaying exponent, a RG analysis shows that
coupling the Luttinger liquids yield a Fermi liquid phase (provided
that the decay exponent is smaller than two again). If all physical
correlators are exponentially decaying (apart from the density one
which always keep, at least, a quadratic decay), the term liquid is often
used. This approach yet remains phenomenological as the
higher-dimension situation and is much more involved.

In this section, we follow the standard practice and consider the
correlation functions of various observables to discuss the nature of
the phases that are realized in the single-mode and two-mode
regimes. The asymptotic decay of the connected correlation functions
associated with the order parameter $\mathcal{O}(x)$ typically reads
$\langle \mathcal{O}(0) \mathcal{O}^\dagger(x)\rangle_c \propto
x^{-\alpha_\mathcal{O}}$ with some exponent $\alpha_\mathcal{O}$. In
order to compute the correlators, we only keep the first harmonics in
Eq.~\eqref{eq:fermions} and begin with the richer case of fermions where
we use the representation in terms of right and left movers:
\begin{equation}
\Psi_{\sigma}(x) \sim e^{i k_{\sigma}x} e^{i(\theta_{\sigma}-\phi_{\sigma})} 
+ e^{-ik_{\sigma}x} e^{i(\theta_{\sigma}+\phi_{\sigma})} \,.
\end{equation}
We use the results that when a field $\phi_a$ is pinned,
$\moy{f(\phi_a)} = f(\moy{\phi_a})$ and its dual $\theta_a$ is
disordered, leading to an exponential decay. In the case of algebraic
correlations, the decay exponents are obtained using the result that,
for a field $\phi$ described by $\mathcal{H}_0$, the equal-time
correlator associated with $\displaystyle A_{m,n}(x) =
e^{i[m\phi(x)+n\theta(x)]}$ behaves asymptotically as
\begin{equation}
  \moy{A_{m,n}(x)A_{-m,-n}(0)} \propto x^{-(m^2 K + n^2/K)/2} \;.
\end{equation}
We now give the leading contributions of the order parameters as a
function of the $a,b$ fields, assuming general
transformation coefficients of the $\up$ and $\down$ modes~:
\begin{widetext}
\begin{align}
\label{eq:green}
\Psi_\sigma(x) & \sim e^{i k_{\sigma}x} e^{-i [\mathfrak{p}_{a\sigma}\phi_a + 
\mathfrak{p}_{b\sigma}\phi_b - \mathfrak{t}_{a\sigma}\theta_{a} -
 \mathfrak{t}_{b\sigma}\theta_{b}]} & \text{Green's function}\\
\label{eq:density}
\hat{n}_{\sigma}(x) & \sim -\mathfrak{p}_{a\sigma}\nabla\phi_a -
 \mathfrak{p}_{b\sigma}\nabla\phi_b + \Lambda^{-1}\cos(2k_{\sigma}x-2(\mathfrak{p}_{a\sigma}\phi_a 
+ \mathfrak{p}_{b\sigma}\phi_b)) & \text{density}\\
\label{eq:singlet}
\Psi_\up(x) \Psi_\down(x) & \sim e^{i(k_{\up}-k_{\down})x} e^{-i[(\mathfrak{p}_{a\down}-
\mathfrak{p}_{a\up})\phi_a + (\mathfrak{p}_{b\down}-\mathfrak{p}_{b\up})\phi_b+
 (\mathfrak{t}_{a\down}+\mathfrak{t}_{a\up})\theta_{a}+(\mathfrak{t}_{b\down}+
\mathfrak{t}_{b\up})\theta_{b}]} & \text{singlet pairing}\\
\label{eq:triplet}
\Psi_\sigma(x) \Psi_\sigma(x) & \sim e^{2i[\mathfrak{t}_{a\sigma}\theta_{a}+
\mathfrak{t}_{b\sigma}\theta_{b}]} &\text{triplet pairing}
\end{align}
\end{widetext}
where $\Lambda$ is a short-range cutoff. Among the multiple
combinations of right and left movers, we have chosen the ones which
should lead to the lowest decay exponents, by having the lowest $m$
and $n$ constant. They usually correspond to the smallest wave-vector.

\subsubsection{The two-mode Luttinger liquid (2M-LL) phase}

In the two mode-regime, all correlators are algebraic
and the leading one will strongly depend on the actual coefficients of
the transformation. The expression of
Eqs.~\eqref{eq:green}--\eqref{eq:triplet} are here understood with general
transformation coefficients of the $\up$ and $\down$ modes as one does
not necessarily have to impose the restriction \eqref{eq:tas} since
$\phi_a$ does not here identify with \eqref{eq:phia}. The transformation
coefficients can be computed exactly~\cite{Mathey2004, Mathey2007a} 
in the absence of the cosine term \eqref{H2pq}. The correlation functions can as well be computed
directly using a Green's function approach~\cite{Orignac2010}.
In the presence of \eqref{H2pq}, the coefficients will be renormalized 
in this two-mode phase to unknown values. This regime is rather generic
and, depending on the interaction and densities,
with many competing orders among which are a Fermi liquid-like phase, a superconducting singlet
or triplet FFLO phase~\cite{Fulde1964} (pairing correlations
displaying the typical $k_{\down}-k_{\up}$), a
spin-density wave (SDW) or charge-density wave (CDW) phase. The case
of equal densities, $p = q = 1$ has the dominant
channels~\cite{Giamarchi2004} among the superconducting, CDW, and SDW
fluctuations. In the cases where spin and charge degrees of freedom
separate, CDW and SDW states are mutually exclusive. Furthermore, for
SU(2)-symmetric models, $x$-, $y$- and $z$-components of the SDW order
parameter are degenerate. These last remarks are no longer valid in
our situation.

\subsubsection{The single-mode Luttinger (1M-LL) multimer phase}

Another regime corresponds to the case where the cosine in Eq.~\eqref{H2pq} 
is relevant in the RG sense. Then, the system has a
massive mode $\phi_a$ given by \eqref{eq:phia}, and a massless mode $\phi_b$.
The massless mode is described in the
low-energy limit by a free bosonic with a velocity $v_{b}$ and a
Luttinger parameter $K_{b}$. In this single-mode
Luttinger liquid, algebraic decays will be ruled by this $K_b$
Luttinger parameter when they occur. When the parameters of the problem satisfy
Eq.~\eqref{eq:condition}, then the massless mode can be found
explicitly. In this section we use these results to discuss 
in details the behavior of the correlation functions.

When $\phi_a$ gets pinned, we see that the above
correlators \eqref{eq:green}--\eqref{eq:triplet} are all exponential
because the presence of $\theta_a$ in their expression,
with the exception of the density one. In particular, all two-body pairing
channels are suppressed, even in the presence of attractive
interactions. In order to construct an operator which has algebraic
correlations, the prefactor in front of $\theta_a$ must vanish. This
is realized by taking the $(p+q)$-mer combination $\Psi_\up^q(x)
\Psi_\down^p(x)$ (bound states of $p$ $\down$-fermions with $q$
$\up$-fermions) which has the prefactor $q\mathfrak{t}_{a\up} +
p\mathfrak{t}_{a\down}$ which is clearly zero from \eqref{eq:tas}:
\begin{widetext}
\begin{align}
\label{eq:pqmer}
\Psi_\up^q(x) \Psi_\down^p(x) & \sim e^{iQ_{qp}x} e^{i[(q\mathfrak{t}_{a\up}
 + p\mathfrak{t}_{a\down})\theta_{a}+(q\mathfrak{t}_{b\up} + p\mathfrak{t}_{b\down})\theta_{b}
 - (s_q\mathfrak{p}_{a\up}-s_p\mathfrak{p}_{a\down})\phi_a - (s_q\mathfrak{p}_{b\up}-
s_p\mathfrak{p}_{b\down})\phi_b]} \;,\\
\intertext{and, in the special case of trimers,}
\label{eq:trimer}
\Psi_\up(x)\Psi_\down(x)\Psi_\down(x) & \sim e^{ik_{\up}x} e^{i[(\mathfrak{t}_{a\up}+
2\mathfrak{t}_{a\down})\theta_{a} + (\mathfrak{t}_{b\up}+2\mathfrak{t}_{b\down})\theta_{b} 
- \mathfrak{p}_{a\up}\phi_a - \mathfrak{p}_{b\up}\phi_b]}\;,
\end{align}
\end{widetext}
in which $Q_{qp} = s_qk_\up - s_pk_\down$ and $s_p = p,
p-2,\ldots,(\text{0 or 1})$, $s_q = q,q-2,\ldots,(\text{0 or 1})$ are
integers accounting for the combination of left and right movers. We
have used a somewhat symbolic notation: by $\Psi^p(x)$, we mean
$\Psi(x+\delta_1)\Psi_(x+\delta_2)\cdots \Psi(x+\delta_p)$, where
$|\delta_i| < \Lambda$ where $\Lambda$ is the short-range cutoff. We stress that
the family of operators \eqref{eq:pqmer} is different from the
``polaronic'' operators introduced in Ref.~\onlinecite{Mathey2004}:
the latter are constructed specifically for minimizing the decay
exponents in the massless phase of \eqref{fullmodel}. On the contrary,
the family \eqref{eq:pqmer} arises naturally in the massive phase of
Eq.\ \eqref{fullmodel}, as a many-body consequence of a existence of
$(p+q)$-body bound states in the microscopic counterpart of
\eqref{fullmodel}.

The effective theory of this $(p+q)$-mer object is then governed by
the gapless mode $b$. Remarkably, as $q\mathfrak{t}_{b\up} +
p\mathfrak{t}_{b\down} = q/\mathfrak{p}_{b\up} = \sqrt{2}\mathcal{K}$,
the exponent is parametrized only by $K_b$, $\mathcal{K}$ and
$p,q$. In order to have the smallest exponent, we have to select the
combination $(s_q,s_p)$ which minimizes the coefficient in front of
$\phi_b$ (one cannot have the combination $p\mathfrak{p}_{b\up} -
q\mathfrak{p}_{b\down}=0$) and which is proportional to $C_{qp} = s_pp
- s_qq$. We list below the coefficients and corresponding wave-vectors
for the simplest commensurabilities:\\
\begin{center}
\begin{tabular}{|c|c|c|c|}
\hline\hline
$(q,p)$ & $(s_q,s_p)$ & $C_{qp}$ & $Q_{qp}$ \\
\hline\hline
$(1,2)$ & $(1,0)$     &  1            & $k_\up$  \\
$(1,3)$ & $(1,1)$     &  2            & $k_\up-k_\down$  \\
$(1,4)$ & $(1,0)$     &  1            & $k_\up$  \\
$(1,5)$ & $(1,1)$     &  4            & $k_\up-k_\down$  \\
$(3,2)$ & $(1,1)$     &  1            & $k_\up$  \\
$(3,4)$ & $(1,1)$     &  1            & $k_\up$  \\
$(3,5)$ & $(1,1)$     &  2            & $k_\up-k_\down$  \\
$(3,7)$ & $(3,1)$     &  2            & $3k_\up-k_\down$  \\
$(5,7)$ & $(1,1)$     &  2            & $k_\up-k_\down$  \\
\hline\hline
\end{tabular}\\
\end{center}
The exponent of the propagator of the $(p+q)$-mer then reads
$\displaystyle \frac 1 2 \left(K_{\text{eff}}^{-1} + C_{qp}^2
  K_{\text{eff}} \right)$ with the effective Luttinger parameter
\begin{equation}
\label{eq:Keff}
K_{\text{eff}} = \frac{K_b}{2\mathcal{K}^2}\;.
\end{equation}
In this phase, the connected density correlations $\mathcal{N}_{\sigma
\sigma'}(x)=\langle n_\sigma(0) n_{\sigma'}(x) \rangle - \langle
n_\sigma(0) \rangle \langle n_{\sigma'}(x) \rangle$ remain algebraic
with the following dominant contributions~:
\begin{gather}
\mathcal{N}_{\up\up}(x) = -\frac{K_\text{eff}}{2 \pi^2}\frac{q^2}{x^2} + 
A_{\up\up}\frac{\cos(2k_\up x)}{x^{2q^2K_\text{eff}}} \;, \label{eq:nupnup} \\
\mathcal{N}_{\down\down}(x) = -\frac{K_\text{eff}}{2 \pi^2}\frac{p^2}{x^2} 
+ A_{\down\down}\frac{\cos(2k_\down x)}{x^{2p^2K_\text{eff}}}\;,\label{eq:dndn}
\end{gather}
where $A_{\sigma\sigma'}$ are non-universal amplitudes. The main
remarks are that (i) the ratio of the zero-momentum fluctuations is
exactly $(q/p)^2$ while the ratio of the density is $q/p$ and (ii) the
wave-vectors are different since $k_\up = \pi n\frac{q}{p+q}$ and
$k_\down = \pi n\frac{p}{p+q}$ as well as their exponents which ratio
should be $(q/p)^2$ exactly. Notice that for the sine-Gordon model,
the ratio of the amplitudes $A_{\up\up} / A_{\down\down}$ are
\emph{exponentially} small in $p-q$ \cite{Lukyanov1997}.

When $C_{qp}=1$, we see that the multimer is effectively behaving as
a spinless fermion (as expected from the combination of a total odd
number of fermions) which Fermi level is $k_{\up}$ and Luttinger
exponent $K_{\text{eff}}$. For instance, trimers belong to this
ensemble. The effective interaction between these spinless fermions,
which are spatially extended objects, is highly non-trivial and
certainly depends on the distance, density and microscopic
parameters (a discussion of such interactions in the case of a boson
mixture can be found in Ref.~\onlinecite{Soyler2009}). However, its
overall effect can be captured by $K_{\text{eff}}$ with effective
repulsion expected when $K_{\text{eff}}<1$ (dominant
CDW fluctuations), and effective attraction expected if
$K_{\text{eff}}>1$ (dominant trimer-pairing
fluctuations). The latter turns out to be a superfluid phase of trimers. 
By associating an even total number of fermions, one should
effectively expect build a bosonic-like multimer. Yet, we see that, in
the propagator of the multimer, one cannot suppress the contribution
from the $\phi_b$ field (as $C_{qp} \neq 0$) and the exponent is not
simply $1/2K_{\text{eff}}$ and thus not simply related to the one of
the density correlations as one would get for a simple bosonic propagator. 
Furthermore, while the momentum distribution
of a boson would usually have a peak at zero-momentum, we see that this
observable will be here diverging at $Q_{qp} \neq 0$.

\subsubsection{The case of a bosonic mixture in the single-mode phase}

As previously mentioned, the effective theory under study can be as
well applied to the situation where the particles are bosons. In the
single-mode phase, a bosonic multimer phase will emerge under the
mode-coupling mechanism and the motivation of this small section is to
discuss the form of the corresponding correlators. We assume repulsive
interactions for the intra-species channels (for stability reasons and
also to lower the $K_{\sigma}$ to be able to fulfill the $K_a<1$
requirement) but attractive interactions in the inter-species channel
(as for the fermions). The boson creator operators are bosonized as
$b_{\sigma} \sim e^{i\theta_\sigma}$ (dropping the higher harmonics
term of Eq.~\eqref{eq:bosons}) which immediately yields
\begin{equation*}
b_\up^q(x) b_\down^p(x) \sim e^{i[(q\mathfrak{t}_{a\up} + p\mathfrak{t}_{a\down})\theta_{a}+(q\mathfrak{t}_{b\up} + p\mathfrak{t}_{b\down})\theta_{b}]}\;.
\end{equation*}
The $(p+q)$-mer is then a true bosonic molecule with an effective
Luttinger parameter which is exactly given by \eqref{eq:Keff}. The
density correlations do not depend on the statistics and still have
the form of \eqref{eq:nupnup}--\eqref{eq:dndn}.

\subsection{Lattice commensurability effects}
\label{sec:lattice}

So far, we have only considered two-component fluids in the continuum
limit which is expected at generic densities on a lattice or in
continuum space. In this section, we briefly discuss the additional
effects arising from the presence of a lattice~\footnote{We assume that
  the field theory description is appropriate --- for too strong
  interactions and/or too large asymmetry it breaks down and the
  lattice model falls into the Falicov-Kimball universality class, see
  Sec~\ref{sec:trimers}.}. An underlying lattice with period $a_0$ can
be viewed as a periodic external potential, in which particles have a
momentum being only defined modulo the reciprocal lattice vector
$2\pi/a_0$. Therefore, umklapp processes with momentum transfer of a
multiple of $2\pi/a_0$ are allowed at low energy. If a Fermi momentum $k_\sigma$
of a species $\sigma$, is itself a multiple of $2\pi/a_0$,
i.e. if a density of species $\sigma$ is commensurate with the
lattice, $s\, n_\sigma = \text{integer}/a_0$, with an integer $s$, an
additional term $\cos(2s\phi_\sigma)$ appears in the low-energy
Hamiltonian. The effects stemming from such a cosine operator
\textit{alone} are well known~: for $K_\sigma \leqslant 2/s^2$ the
cosine is relevant in the RG sense and the system undergoes a Mott
transition into a density wave state with the unit cell of $s$ lattice
sites. In a two-component system, it is possible to have two operators
of this sort, one for each species. Furthermore, if the densities are
such that $s n_\up + s' n_\down$ is an integer (we set $a_0=1$ from
now on) for some integers $s$ and $s'$, there is yet another term in
the low-energy Hamiltonian, namely $\cos{ 2(s\phi_\up + s'\phi_\down)
}$ (cf. Eq.~\eqref{H2}).

Here, we analyze a simple special case where
\begin{align}
p\,n_\up - q\, n_\down &= 0 \;, \label{latt:pqrl1}\\
r\,n_\up + l\, n_\down &= 1 \; ,\label{latt:pqrl2}
\end{align}
or $n_\up = q/(pl+qr)$ and $n_\down = p/(pl+qr)$ with the integers
$p$, $q$, $r$ and $l$. Given \eqref{latt:pqrl1}--\eqref{latt:pqrl2},
Eq.~\eqref{HaldaneN} yields the Hamiltonian in the form
$\mathcal{H}_0(\phi_\up) +\mathcal{H}_0(\phi_\down) + \mathcal{H}_1+
\mathcal{H}_\mathrm{cos}$ with
\begin{eqnarray}
\mathcal{H}_\mathrm{cos} =&& G_1 \int\! dx \, \cos{ 2(p\,\phi_\up - q\,\phi_\down) } \label{latt:H1} \\
                         &+& G_2 \int\! dx \, \cos{ 2(r\,\phi_\up + l\,\phi_\down) } \label{latt:H2} \\
                         &+& G_3 \int\! dx \, \cos{ 2(pl+qr)\,\phi_\down } \label{latt:H3}  \\
                         &+& G_4 \int\! dx \, \cos{ 2(pl+qr)\,\phi_\up }  \label{latt:H4} \; .
\end{eqnarray}
where $G_1$,\dots,$G_4$ are non-universal amplitudes. Interpretation
of Eqs.\ \eqref{latt:H1}--\eqref{latt:H4} is straightforward: Eq.\
\eqref{latt:H1} stems from the condition \eqref{latt:pqrl1} and is
thus insentive to the presence of the lattice (cf. Sec.\
\ref{sec:modecoupling}); Eqs.\ \eqref{latt:H3} and \eqref{latt:H4}
favor the Mott localization of the species $\down$ and $\up$,
respectively. On the other hand, operator \eqref{latt:H2} is unique to
two-component lattice systems and owes its existence to the peculiar
commensurability condition \eqref{latt:pqrl2}. The physical meaning of
\eqref{latt:pqrl2} is clear: by analogy with Sec.\ \ref{subsec:corr},
it favors the quasi long-range ordering of the operator
$\mathcal{O}_{r+l} = \Psi_\down^r(\Psi_\up^\dagger)^l$.

\begin{figure}[t]
\centering
\includegraphics[width=\columnwidth,keepaspectratio=true,clip]{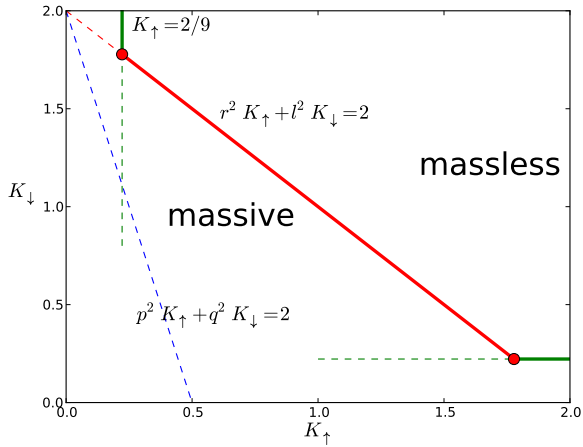}
\caption{(Color online). Diagram showing the effect of commensurate
densities (see text for discussion) in the special case of
$n_\up = 1/3$, $n_\down=2/3$, i.e. $p=2$ and
$q=r=l=1$. For $2-2/9 < K_\down < 2/9$ the interaction with the
lattice leads to a formation of a 'trimer crystal' state. For larger(smaller)
values of $K_\down$ the system has a phase transition from a
massless phase into a Mott insulator of $\up$($\down$)-component.
The 'trimer' operator $\cos{2(2\phi_\up-\phi_\down)}$ is always
subdominant.}
\label{latt:fig_tommaso}
\end{figure}

\begin{figure}[t]
\includegraphics[width=\columnwidth,keepaspectratio=true,clip]{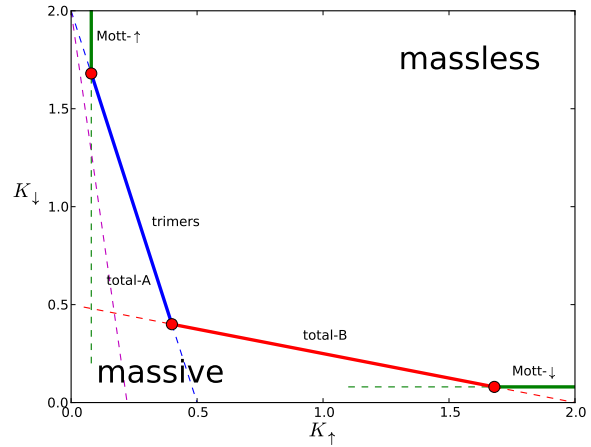}
\caption{(Color online). Same as Fig.~\ref{latt:fig_tommaso} for
$n_\up = 1/5$, $n_\down=2/5$. In this case, Eqs.\
\eqref{latt:pqrl1}--\eqref{latt:pqrl2} allow \emph{two} sets of
solutions: (A) $l=1$ and $r=3$, and (B) $l=2$ and $r=1$, with $p=2$
and $q=1$ in both cases. Solution (A) is always subdominant, while
(B) dominates in the window $2/5<K_\down<2/25$. For
$48/25<K_\down<2/5$, the dominant instability is the formation of a
Luttinger liquid of trimers.}
\label{latt:fig_nup15ndown25}
\end{figure}

In the following, for sake of simplicity, we assume equal velocities
of the two-components and drop the $\mathcal{H}_1$ term.
The dominant instability of the massless theory
$\mathcal{H}_0(\phi_\up) + \mathcal{H}_0(\phi_\down)$ is due to the
operator with largest positive scaling dimension. Depending on the
values of $K_\up$ and $K_\down$, the following inequalities define
which of the operators \eqref{latt:H1}--\eqref{latt:H4} is relevant:
\begin{align}
p^2 K_\up + q^2 K_\down &\leqslant 2 \label{latt:dim_pq} \; , \\
r^2 K_\up + l^2 K_\down &\leqslant 2 \label{latt:dim_rl} \; , \\
(pl+qr)^2 K_\down & \leqslant 2 \label{latt:dim_down} \;, \\
(pl+qr)^2 K_\up & \leqslant 2 \label{latt:dim_up} \;, 
\end{align}
respectively. In Figs.\ \ref{latt:fig_tommaso} and
\ref{latt:fig_nup15ndown25} we plot the $(K_\up,K_\down)$ diagrams
corresponding to Eqs.\ \eqref{latt:dim_pq}--\eqref{latt:dim_up} for
two values of the densities.  We see that which instability takes
place depends on the values of the bare Luttinger parameters $K_\up$
and $K_\down$, and thus on microscopic details of an underlying
lattice model. Numerically, a crystal phase has been reported~\cite{Keilmann2009} 
in a two-component bosonic Hubbard model and a similar result
is presented in the fermionic counterpart in Sec.~\ref{sec:crystal}
for the commensurabilities discussed in Fig.~\ref{latt:fig_tommaso}.
These phases do correspond to the locking of several combinations of 
the modes according to Eqs.\ \eqref{latt:H1}--\eqref{latt:H4} but
they are achieved for very large asymmetry. Consequently, 
the above criteria \eqref{latt:dim_pq}--\eqref{latt:dim_up} determined 
for equal velocities are not directly applicable in these situations. 
The quantitative predictions of \eqref{latt:dim_pq}--\eqref{latt:dim_up}
could be relevant to the case of strongly renormalized $K_{\sigma}$,
for instance with long-range intra-species interactions.

A striking feature of the phase diagrams \ref{latt:fig_tommaso} and
\ref{latt:fig_nup15ndown25} is the appearance of the multicritical
points where several instabilities compete.  In the above treatment we
have only considered an effect of various operators
\eqref{latt:H1}--\eqref{latt:H4} \emph{alone}. An interplay between
different operators is non-trivial and may lead to consequences not
captured by the simple power counting of
Eqs.~\eqref{latt:dim_pq}--\eqref{latt:dim_up}. Hence, applicability of
the above analysis in the vicinities of the multicritical points is
not granted. There are several possible scenarios of the phase
transitions at such multicritical points. For one thing, it is easy to
construct fine-tuned theories where two continuous transitions occur
simultaneously. An other possibility is a first order transition, as
been observed in numerical simulations of higher-dimensional bosonic
systems \cite{Kuklov2004}. Detailed analysis of these
multicritical points is beyond the scope of the present paper.

\section{Trimer formation in the 1D asymmetric Hubbard model}
\label{sec:trimers}

In this second part, we study the emergence of a trimer phase  
on a particular microscopic model :
the 1D asymmetric attractive Hubbard model. After defining the model and providing
its phase diagram as a function of the parameters, we discuss some
limitations of the bosonization approach to this model and an
alternative phenomenological description that completes
the interpretation of the obtained data.

\subsection{Model and qualitative aspects}

We consider two species of fermions which internal degree of freedom
is denoted by a spin index $\sigma$. They hop on a lattice with a
spin-dependent amplitudes $t_{\sigma}$ (which would experimentally
corresponds to different optical lattices for each species) and
interact locally only in the inter-species channel with a Hubbard term
$U$ which we take negative, as suggested by the arguments of
Sec.~\ref{sec:boso} and as a natural choice to favor bonding between
particles. The Hamiltonian is then :
\begin{equation}
\label{eq:hubbard}
\mathcal{H} = -\sum_{i,\sigma=\up,\down} t_{\sigma} 
[c^{\dag}_{i+1,\sigma}c_{i,\sigma} + \text{h.c.}] + U\sum_i n_{i,\up}n_{i,\down}\;.
\end{equation}
One of the key parameter for the physics is the ratio between the
hoppings $\eta = t_{\down}/t_{\up}$. In order to have the possibility
of forming trimers, we take the commensurate condition $2n_{\up} =
n_{\down}$ but the total density $n$ varies freely and is another
important parameter of the physics. Using the notations of
Sec.~\ref{sec:boso}, we thus have $p=2$ and $q=1$ (the simplest new
combination one can have). The Fermi momenta are $k_{\sigma} = \pi
n_{\sigma}$ and free fermions Fermi velocities read $v_{\sigma} =
2t_{\sigma} \sin(\pi n_{\sigma})$. Since $n_\down \leq 1$, the maximum
total density one can have for this commensurability is $n=3/2$.

The above Hamiltonian has been widely studied in the case of
balanced~\cite{Giamarchi1995} and imbalanced densities~\cite{Yang2001}
but the special commensurability where trimers emerge
has only been investigated for one set of data in
Ref.~\onlinecite{Burovski2009}, showing that the pairing correlations
were indeed suppressed, in agreement with the bosonization
approach. When the asymmetry is very large, one species behaves
quasi-classically (they get localized) and the model is in the regime
of the Falicov-Kimball (FK) model~\cite{Falicov1969} where there exists
a lot of quasi-degenerate states at low-energies, analogue to a phase
separation regime. We expect generically a first-order transition to
this segregated (or demixed) phase when lowering $\eta$ in the phase
diagrams. The FK regime can display rather rich physics recently
investigated in Ref.~\onlinecite{Barbiero2010} and which will not be
analyzed here : our aim is only to draw the boundary of this regime.
Numerically, the transition to the FK is rather sharp and all
observables clearly display segregation.
For $\eta=1$, the arguments of Sec.~\ref{sec:boso} suggest that the
two-mode regime will be generically realized. Qualitatively, in a
strong-coupling picture where two spin-$\down$ fermions are localized
on neighboring sites, the delocalization of a spin-$\up$ electron on
these sites will be favored by attractive interactions, forming a very
local trimer state. This picture will be correct at small enough
densities and actually not too large $U$ and too small $\eta$,
otherwise such bound states will agglomerate with other spin-$\up$ and
$\down$ fermions, leading to the FK regime. We thus expect the
formation of the trimer phase in the vicinity of the FK but at both
finite $U$ and finite $\eta$. Within the framework of
Sec.~\ref{sec:boso} and considering that the starting point of
bosonization are free fermions, the ratio between the velocities $\nu
= v_\down/v_\up = 2 \eta \cos(\pi n /3)$ supports that small $\eta$
clearly favors the formation of trimers while small densities should
not. 

\begin{figure}[t]
\centering
\includegraphics[width=0.95\columnwidth,clip]{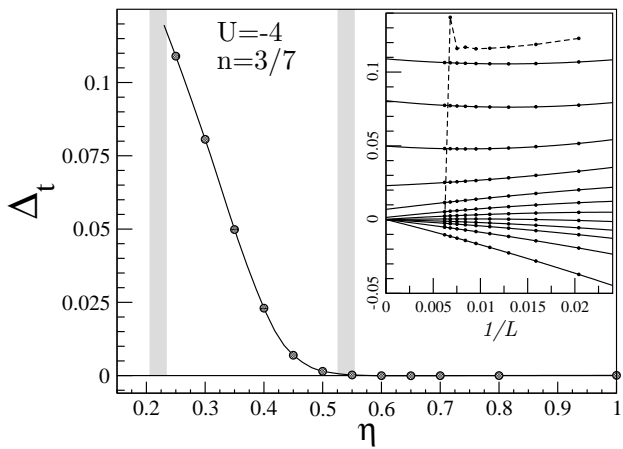}
\caption{Opening of the trimer gap increasing mass asymmetry (lowering
  $\eta= t_\down/t_\up$) for a fixed interaction and density. The
  magnitude of the gap (in units of $t_\up$) is small in comparison to $U$ and $t_\up$. The
  grey areas are estimates of the transition points. \emph{Inset:} finite
  size extrapolations of the gap. The upper dashed curve shows the
  behavior for $\eta=0.2$ when entering in the FK regime.}
\label{fig:gap}
\end{figure}

\subsection{Phase diagrams}

The phase diagrams of model~\eqref{eq:hubbard} are numerically
determined using standard DMRG with
open-boundary conditions (OBC) and keeping up to $M = 2000$ states. In
order to discriminate between the different possible regimes, we use
both ``global'' probes and local observables and correlation
functions. Among ``global'' probes, one can use the trimer gap
$\Delta_{\text{t}}$ associated with the formation of the bound
state. It can be defined following Ref.~\onlinecite{Orso2010} as:
\begin{equation}
\begin{split}
\Delta_{\text{t}} = &E_0(N_\up+1,N_\down+1) + E_0(N_\up,N_\down+1) \\
&- E_0(N_\up+1,N_\down+2) - E_0(N_\up,N_\down)\;,
\end{split}
\end{equation}
with $E_0(N_\up,N_\down)$ the ground-state energy with $N_\up,N_\down$
fermions. Results as the function of the asymmetry $\eta$ for an
incommensurate density $n=3/7$ and large interaction $U=-4t_\up$ have
been extrapolated to the thermodynamical limit and are given in
Fig.~\ref{fig:gap}. The slow opening of the trimer gap is
\emph{qualitatively} compatible with the sine-Gordon behavior of
Sec.~\ref{sec:boso} although the transformation is not directly
applicable for any $\eta$. Notice that the whole system remains
gapless. The slow opening of the gap makes it difficult to precisely
locate the transition point. In such situation, a usual approach would
be to use the prediction on the critical Luttinger parameter
$K_a=K_a^c$ at the transition point. Furthermore, the determination of
$K_a$ using correlators in the two-mode phase is very
difficult as it would require to know, and then to disentangle,
the complicated expression of the exponents as a function of
$K_a$ and $K_b$ to extract them independently.

\begin{figure}[t]
\centering
\includegraphics[width=0.8\columnwidth,clip]{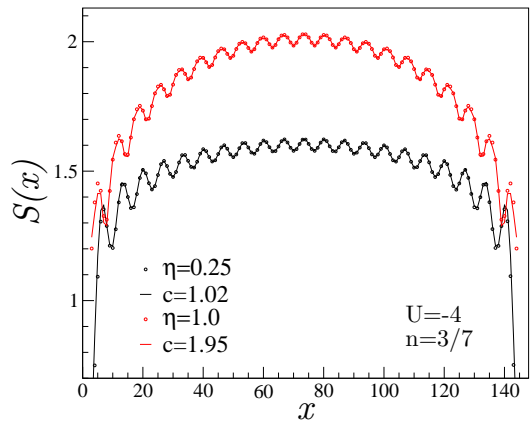}
\caption{(Color online). Examples of fits of the entanglement entropy
  in the two-mode and single-mode phases using Eq.~\eqref{eq:EE}. It
  shows a clear quantitative difference with respectively $c\simeq 2$
  and $c\simeq 1$, as expected.}
\label{fig:EEexamples}
\end{figure}

\begin{figure}[t]
\centering
\includegraphics[width = 0.85\columnwidth,clip]{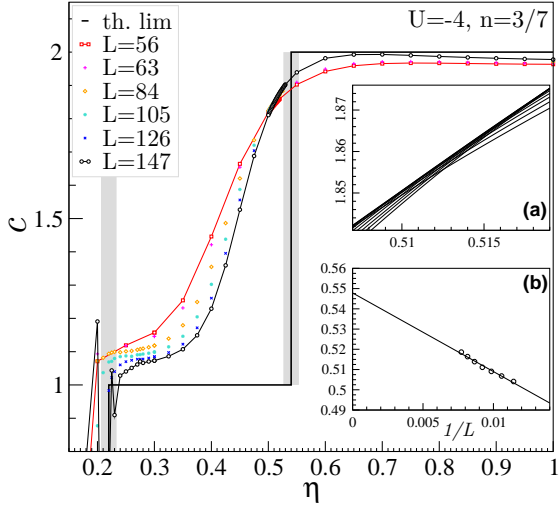}
\caption{(Color online). Central charge $c$ obtained from fits as in
  Fig.~\ref{fig:EEexamples} as a function of the asymmetry $\eta$ for
  $U=-4t_\up$ and $n=3/7$. The stair-like behavior with increasing
  system size $L$ allows an efficient determination of the
  transition. In the Falicov-Kimball regime (left), fits give
  $c\simeq 0$ or irrelevant numbers. \textbf{(a)} Zoom of the
  $c(L)$ curves in the crossing region illustrating the extraction of
  crossing points between successive sizes. \textbf{(b)} Tentative
  finite-size extrapolation of the crossing points $\eta_c(L)$.}
\label{fig:centralchargeU-4}
\end{figure}

Therefore, we use another global approach to the distinction between
the two-mode and single-mode phases, which is particularly well-suited
for this model, and more generally in a similar context. 
Using universal results on the entanglement entropy
(EE), the central charge $c$ of the model can be extracted which
directly gives access to the number of bosonic modes, without further
information on their nature. Hence, we expect $c=2$ in the two-mode
regime while $c=1$ in the single-mode trimer phase. This stair-like
expectation in the thermodynamical limit will be smoothed out by
finite-size effects. The central charge is obtained on finite-systems
using the following ansatz for the EE between a left block of size $x$
and the right block of length $L+1-x$ with OBC:
\begin{equation}
\label{eq:EE}
S(x) = \frac c 6 \ln d(x|L+1) + A\,t(x) + B\;,
\end{equation}
where $d(x|L)$ is the cord function
\begin{equation}
\label{eq:cord}
d(x|L) = \frac L \pi \sin\left(\frac{\pi x}{L}\right)\;,
\end{equation}
and $t(x)$ is the local kinetic energy on bound $(x,x+1)$ (obtained
numerically), and $A,B$ are fitting parameters. The first log term is
the leading and universal one~\cite{Calabrese2004}
while the second accounts for finite-size oscillations due to OBC and
which can have a significant magnitude~\cite{Laflorencie2006, Roux2009}. It is thus essential to take them into
account to improve the quality of the fits. In the end, there are only
three parameters in the procedure and typical examples in both the
two-mode and single-mode phases are given in
Fig.~\ref{fig:EEexamples}. Systematic fits on finite-size systems
provide an estimate of $c$ as a function of the parameters. As seen in
Fig.~\ref{fig:centralchargeU-4}, the $c(L)$ curves crosses around the
transition point. Although we do not have any quantitative prediction
for the finite-size corrections of $c(L)$ obtained in this way, we can
argue that if $L$ is smaller than the correlation length associated
with the trimer gap, $c(L)$ will be larger than one as the system is
effectively in a two-mode regime. Thus, $c(L)$ should decrease with $L$
towards one in the single-mode phase, as observed. In the two-mode
phase, there is no obvious discussion : we only expect that the larger
the system, the better the agreement with the continuous limit. One
can also check the effect of the number of kept states on the fits and
see that they don't have the dominant effect in this model which
converges well numerically. We have estimated the transition point by
extrapolating the crossing points between successive sizes (see
Fig.~\ref{fig:centralchargeU-4}\textbf{(a)}) as a function of the
inverse size (see Fig.~\ref{fig:centralchargeU-4}\textbf{(b)}). From
this approach and the opening of the gap, we get a critical value
$\eta_c \simeq 0.54\pm 0.02$ for the mass asymmetry on this
cut. Although the gap is rather small, the trimer region appears to be
rather wide.

\begin{figure}[t]
\includegraphics[width=\columnwidth,clip]{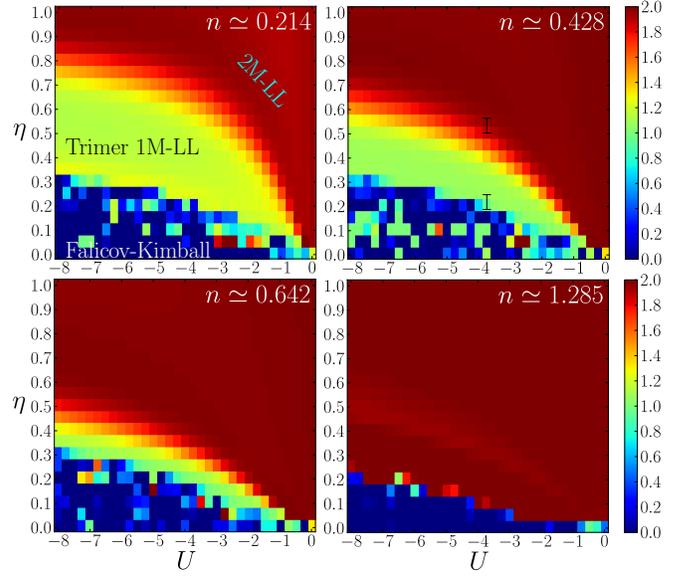}
\caption{(Color online). Maps of the central charge $c$ vs interaction $U$ and asymmetry $\eta$
  for a system with $L=112$ at four different densities. For
  $n=3/7\simeq 0.428$, the lines with error bars are the ones
  estimated from Figs.~\ref{fig:gap}
  and~\ref{fig:centralchargeU-4}. The $\eta=0$ cuts correspond to
  data obtained with a very low but non-zero value $\eta=0.005$.}
\label{fig:PhaseDiagN}
\end{figure}

\begin{figure}[t]
\centering
\includegraphics[width=\columnwidth,clip]{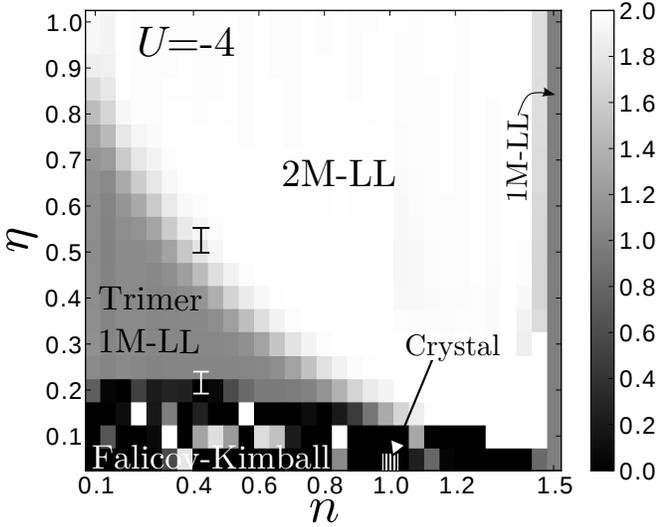}
\caption{Map of the central charge $c$ vs asymmetry $\eta$ and total density $n$ for fixed
  interaction $U=-4t_\up$ on a system with $L=128$. The lines with
  error bars are the ones estimated from Figs.~\ref{fig:gap}
  and~\ref{fig:centralchargeU-4}.}
\label{fig:PhaseDiagU}
\end{figure}

Using the central charge calculation, one can map out the phase
diagram in the $(\eta,U)$ plane for a fixed density, or in the
$(\eta,n)$ plane for a fixed interaction $U$. Results are gathered in
Fig.~\ref{fig:PhaseDiagN} and \ref{fig:PhaseDiagU} respectively. 
These diagrams display bare data for a
given system with a rather large size $L=112$ and the previous
estimate of the cut is given as error bars. These diagrams show that a
wide trimer phase can be achieved at large enough interactions, small
enough $\eta$, as expected, and also that low densities strongly
favors their formation. At large densities $n\simeq 1.3$, the trimer
region vanishes within our grid resolution so that it is at most confined to a
very tiny region between the two-mode phase and the FK regime. While
the large-$|U|$ situation is rather clear, the competition between the
three regimes at small $U$ is more involved. Indeed, two scenarios can
occur in the $(\eta,U)$ plane: either the trimer phase always
separates the FK and two-mode regimes, corresponding to two boundaries
starting from the $(\eta=0,U=0)$ corner, or there is a critical
$\vert{U}\vert$ above which the trimer phase emerges, corresponding to
a tricritical point $(\eta_c,U_c)$. We could not numerically
discriminate between both scenarios, but we do find a small trimer
region at relatively small $U$s ($U\simeq -1\,t_\up,-2\,t_\up$) for most
densities : we do not have evidence for a tricritical point with a
large $U_c$. As the density plays a central role in the stabilization
of the trimer phase, we give in Fig.~\ref{fig:PhaseDiagU} the central
charge map for a fixed interaction $U=-4t_\up$ as a function of the total density
$n$ and mass asymmetry. A similar question about a intervening trimer
phase between the two-mode and the FK regimes can be raised. While the
two-mode and FK are clearly separated at small densities, we found
that if a trimer intermediate phase exists at large densities up to
the $(\eta=0,n=1.5)$ point, its extension will be particularly small
(not seen within our numerical calculations). In addition to
the three main phases, commensurability effects are also present in
this diagram. When the maximum density $n=1.5$ is reached, the
$\down$-band is completely filled while the $\up$-band is half-filled,
leading to a single-mode phase well-captured by the central charge
approach. Lastly, as it will be discussed in Sec.~\ref{sec:crystal}, a
crystal phase (fully gapped) exists for the commensurate density $n=1$
at very small $\eta$ and is pointed on
Fig.~\ref{fig:PhaseDiagU}. Other commensurabilities could yield
additional crystal-like phases in this diagram but this is beyond the
scope of this study.

\subsection{Observables and effective behavior of the trimer liquid}

In this section, we give the behavior of several observables in order
to see how they are affected by the entrance into the trimer phase or
the FK regime and, as well, to investigate the effective behavior of
the trimer fermion.

\begin{figure}[t]
\centering
\includegraphics[width=\columnwidth,clip]{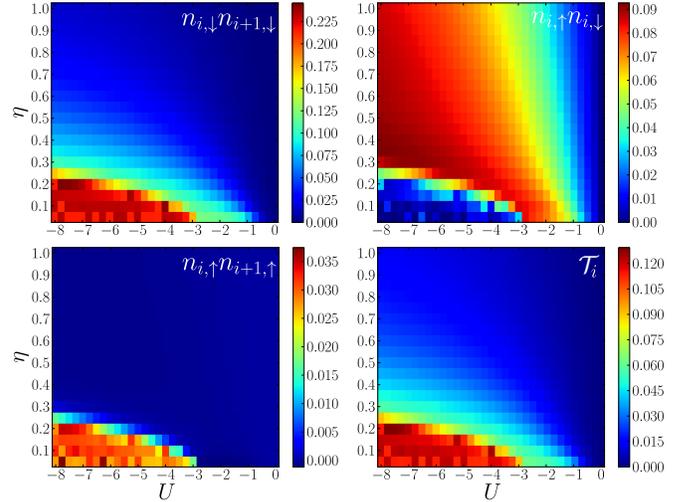}
\caption{(Color online). Maps of averaged local quantities in a system
  with $L=56$ at total density $n=3/7\simeq 0.428$. The non-interacting
  expectation ($U=0$ line) has been subtracted in order to unveil the
  effect of the interaction.}
\label{fig:local_maps}
\end{figure}

\subsubsection{Local observables}

First, we select a set of local correlators
(living on sites or on bonds) which illustrate the phenomenological
picture of the different parts of the phase diagram. We compute the
local double occupancy $\moy{n_{i,\up}n_{i,\down}}$, the trimer local
operator as $\mathcal{T}_i = \frac 1 2
(\moy{n_{i,\up}n_{i,\down}n_{i+1,\down}}+
\moy{n_{i+1,\up}n_{i,\down}n_{i+1,\down}})$ (since the light particle
is in principle delocalized above two heavier), and the density
correlators $\moy{n_{i,\up}n_{i+1,\up}}$ and
$\moy{n_{i,\down}n_{i+1,\down}}$. This choice of local correlators is
well suited to a strong-coupling picture as pairs or trimers should in
principle correspond to a narrow bound-state, spread over only a few
lattice sites. These local correlators should then pick up a
reasonable weight of the local bound-state. The results are averaged
over all lattice sites and plotted in Fig.~\ref{fig:local_maps}. The
expectation value at $U=0$ has been subtracted so that the reference
state is the free fermions limit at a given $\eta$ (the pairing or trimer local
correlators defined above are obviously non-zero even in the free
fermions limit). Fig.~\ref{fig:local_maps} display behaviors in
qualitative agreement with the picture we have on the trimer formation : the
${\up\up}$ density correlator is nearly zero everywhere but in the FK
regime, signaling phase separation. On the contrary ${\down\down}$
density correlator increases significantly in the region corresponding
to the trimer phase, surrounding the FK pocket, and together with the
local trimer density $\mathcal{T}_i$. Lastly, we see that the double
occupancies, or pairs, acquire a strong weight with negative $U$
everywhere in two-mode and single-mode regions : they are either
``independent'' or embedded in the trimer bound-state. Their coherence
can yet be probed only by measuring correlations as discussed below.

\begin{figure}[t]
\centering
\includegraphics[width=0.9\columnwidth,clip]{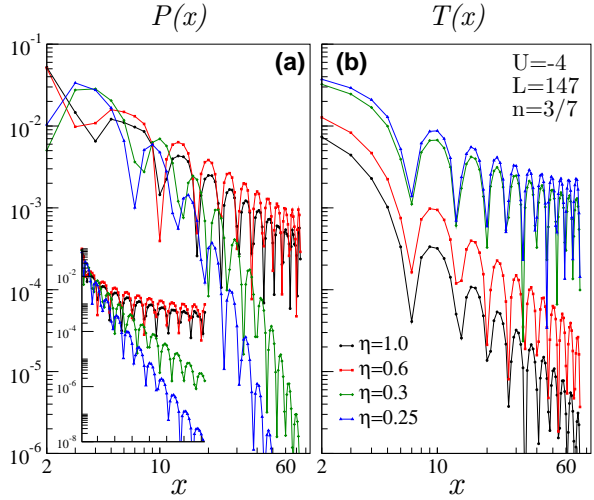}
\caption{(Color online). Behavior of \textbf{(a)} pairing and
  \textbf{(b)} trimer correlations when lowering $\eta$ and
  entering the trimer phase along a cut at $U=-4t_\up$ in the phase
  diagram (absolute values are displayed). The Inset of \textbf{(a)}
  shows the same data but in log-linear scale to highlight the
  exponential decay.}
\label{fig:TPcorrelations}
\end{figure}

\subsubsection{Pairing and trimer correlations, effective behavior of the trimers}

\begin{figure}[t]
\centering
\includegraphics[width=0.9\columnwidth,clip]{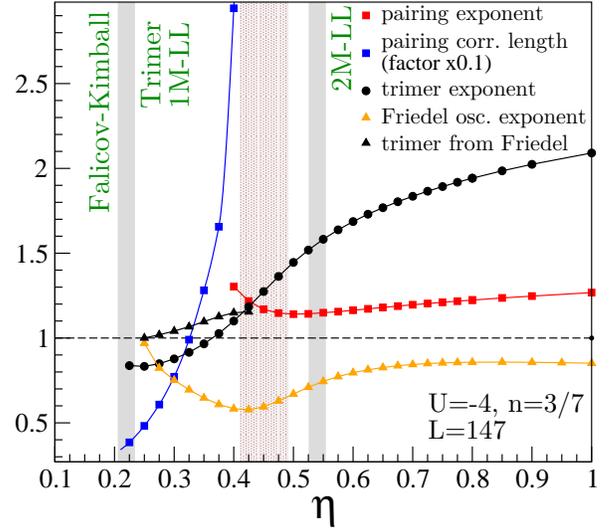}
\caption{(Color online). Comparison of the decay exponents of pairing
  and trimer correlations. The correlation length of pairing
  correlation (in units of the lattice spacing) is given in the trimer
  phase. The strong grey areas are the previous estimates of the
  transition points while the light maroon area illustrates the
  location of the transition on the $L=147$ finite-size system under
  study. The exponent of the Friedel oscillations of $n_\up$ is also
  displayed, together with the expected trimer exponent derived from it
   (see text for discussion).}
\label{fig:exponent}
\end{figure}

\begin{figure}[t]
\centering
\includegraphics[width=\columnwidth,clip]{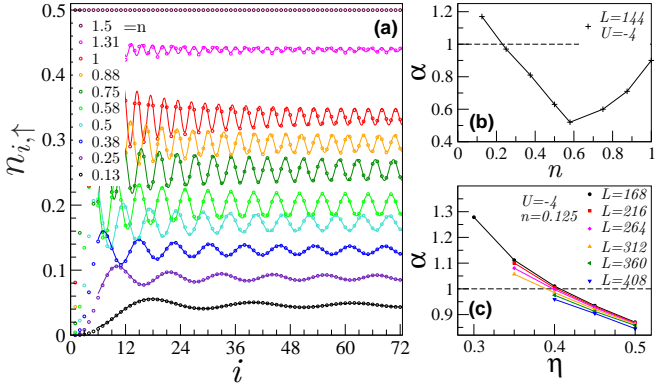}
\caption{(Color online). \textbf{(a)} Typical Friedel oscillations of
  the $\up$-density for $U=-4t_\up$, $\eta=0.3$ and $L=144$ for
  various densities $n$. Full lines are fits using
  Eq.~\eqref{eq:friedel}. \textbf{(b)} Decay exponents obtained from
  the fits as a function of the density. The cusp at $n \simeq 0.6$
  roughly corresponds to the transition from the two-mode to the
  single-mode regime. \textbf{(c)} Large exponents at low densities,
  close to the FK regime when lowering $\eta$ : increasing the size
  tends to reduce the exponent below one.}
\label{fig:density}
\end{figure}

We now turn to the behavior of correlation functions across the phase
diagram. From Sec.~\ref{subsec:corr}, and as already observed for a
particular point in Refs.~\onlinecite{Burovski2009, Orso2010}, the
pairing correlations change from algebraic to exponential decay when
entering into the trimer phase. These correlations are here computed
in the singlet channel and for local pairs $\hat{P}_i =
c_{i,\up}c_{i,\down}$. In addition, we compute the trimer correlator
using the local trimer operator $\hat{T}_i =
c_{i,\up}c_{i,\down}c_{i+1,\down}$ defined on neighboring sites. The
associated correlation functions $P(x) =
\moy{\hat{P}^\dag_i\hat{P}_{i+x}}$ and $T(x) =
\moy{\hat{T}^{\dag}_i\hat{T}_{i+x}}$ are computed with $i$ taken at
the center of the chain. Increasing the mass asymmetry along the same
cut at $U=-4t_\up$ as in previous figures, the suppression of pairing correlations is
clearly seen in Fig.~\ref{fig:TPcorrelations}\textbf{(a)}. On the
contrary, trimer correlations, which are subdominant in the two-mode
regime, are boosted by smaller $\eta$, both in amplitude (as for the
local correlators previously evoked) and in the decay exponent which
gets smaller (see Fig.~\ref{fig:TPcorrelations}\textbf{(b)}). Notice
that the wave-vector is the same for both correlators since $k_\down -
k_\up=k_\up=\pi n /3$ for this commensurability. We have tentatively
extracted the decay exponents of both correlators by fitting the
functions using a power-law modulated by a cosine oscillations. The
correlation length $\xi$ of the pairing correlator in the trimer phase
is obtained using an exponential envelope $e^{-x/\xi}$. The results are gathered in
Fig.~\ref{fig:exponent}, showing the evolution in both phases. We must stress
that the data computed on a finite-size system display a transition at
a lower $\eta$ than in the thermodynamical limit. From
Sec.~\ref{subsec:corr}, we expect that the decay exponent of the
trimer propagator is of the form
$(K_{\text{eff}}+K_{\text{eff}}^{-1})/2$ while the $\up$-density
correlations have a decay exponent of $2K_{\text{eff}}$. A first
consequence is that the trimer exponent should always be larger than
one which is not reproduced for the lowest $\eta$s, and which we
attribute to numerical inaccuracies of the fits of trimer correlations. 
Besides, inverting
$(K_{\text{eff}}+K_{\text{eff}}^{-1})/2$ to get $K_{\text{eff}}$ is
subjected to strong errors when $K_{\text{eff}}\simeq 1$ and does not
tell whether $K_{\text{eff}}>1$ or $K_{\text{eff}}<1$ which is
essential for the effective behavior of the trimers. In order to get
a better estimate of $K_{\text{eff}}$, we rather use Friedel oscillations of
the $\up$-density operator which decay exponent $\alpha$ is equal to
$K_{\text{eff}}$ in the trimer phase according to Sec.~\ref{subsec:corr}.
Even though the approach of Sec.~\ref{subsec:corr} is not
applicable for most parameters, the fact that there exists an
effective Luttinger exponent $K_{\text{eff}}$ describing the physics
of the fermionic trimer with a propagator
$(K_{\text{eff}}+K_{\text{eff}}^{-1})/2$ and with Friedel
oscillations with $K_{\text{eff}}$ is more general : the limitation of
the bosonization approach is rather that $K_{\text{eff}}$ will not take the form of
Eq.~\eqref{eq:Keff}. Local observables are believed to have less
numerical errors associated with a finite number of kept states than
correlations~\cite{White2007}. Thus, we fit the Friedel oscillations
of the $\up$-density using the following symmetric ansatz
\begin{equation}
\label{eq:friedel}
n_{i,\uparrow} = n_0 + A\frac{\cos\,q(i -\frac{L+1}{2})}{[d(i|L+1)]^{\alpha}}
\end{equation}
with $1 \leq x \leq L$ and only four fitting parameters $n_0$, $A$,
$q$ and $\alpha$~\footnote{We expect that $q = 2\pi n_{\uparrow} = 2\pi n/3$
and $n_0 = n_{\uparrow}$ but at low densities, it is usually better to take $q$
and $n_0$ as free independent fitting parameters due to the depletion
of the density at the edges which effectively increases it in the
bulk. For instance, one has $n_0=(N+1/2)/(L+1)$ for free spinless
fermions on a finite system with $N$ fermions, which is not exactly
the average density $N/L$, particularly at small $N/L$.}.
In the trimer phase, we expect $\alpha = K_{\text{eff}}$. Some typical fits are
plotted in Fig.~\ref{fig:density}\textbf{(a)}. From them, we
extract the decay exponent $\alpha$ and plot it on
Fig.~\ref{fig:density}\textbf{(b)} as a function of the total
density. A cusp is found around $n\simeq 0.6$ signaling the transition
from the two-mode regime to the trimer phase. We have seen that a low
density favors the formation of the trimer phase. This figure shows
that, in the trimer
phase, we have $K_{\text{eff}}<1$  for the larger densities,
corresponding to a repulsive effective interaction between trimers
(dominant CDW order of trimers). We observe that the
exponent increases with decreasing density, compatible with the fact
that at low densities in the trimer regime, the trimers should be
close to free spinless fermions having $K_{\text{eff}}\simeq 1$. 
Bare data for the smallest density on a system with $L=144$ even
display an exponent $K_{\text{eff}}\simeq 1.2$ larger than one. 
Interestingly, the trimers in this model are necessarily objects
with a finite extension of at least two sites and two close trimers
may have the possibility to overlap by delocalizing their $\up$
electrons. The distance dependence and sign of the effective
interaction between trimers is non-trivial -- a perturbation
theory to derive it looks challenging as it
involves many sites and degrees of freedom. Yet, since the
trimer phase is found close to the FK regime, we can expect the effective interaction to become
attractive close to this boundary, leading to $K_{\text{eff}}>1$. Such
physics would correspond to a superfluid phase of trimers. This would
be physically very remarkable since the microscopic
Hamiltonian~\eqref{eq:hubbard} would contain both the formation of
bound-states, or molecules, and their effective superfluid
behavior. However, the behavior close to a phase separation at small
densities is numerically involved. Indeed, increasing the system size
$L$ shows that $\alpha$ actually tends to decrease below one, as
reported in Fig.~\ref{fig:density}\textbf{(c)}, or one enters the FK
regime for larger sizes. We did not find clear evidence of a
stabilization of $K_{\text{eff}}>1$ in the thermodynamical limit and
understand the observed $K_{\text{eff}}>1$ as finite size effects. A
superfluid droplet picture can be qualitatively put forward. Starting
from the FK regime and looking at the local density pattern, one sees
that the fermions are clustered into droplets while other parts of the
box are empty. Approaching the trimer phase from the FK regime tends
to increase the size of these droplets to gain kinetic energy. When a
box confinement is present (finite system with OBC), it naturally favors the
overlap between trimers, by depleting the edges, and can prevent the
droplet to form (for instance if their typical size is larger than the
box size). Increasing further the size of the box (at constant
density) can lead to droplets formation. 
This is a possible interpretation of the data
observed in Fig.~\ref{fig:density}\textbf{(c)}.
In addition, we must stress
that there is a lot of competing low-energy states in the FK regimes
so DMRG, as an essentially variational methods, can be trapped into
metastable states. Even though the
thermodynamical limit is unclear, it is experimentally motivating to
have signatures of superfluidity on mesoscopic confined systems as one
has in cold-atoms setups. We further mention that a recent careful
study of the t-J model on a chain~\cite{Moreno2010} which could
qualitatively contain a similar phenomenon as pair-clustering did not
find evidences for such clustering.

\begin{figure}[t]
\centering
\includegraphics[width=\columnwidth,clip]{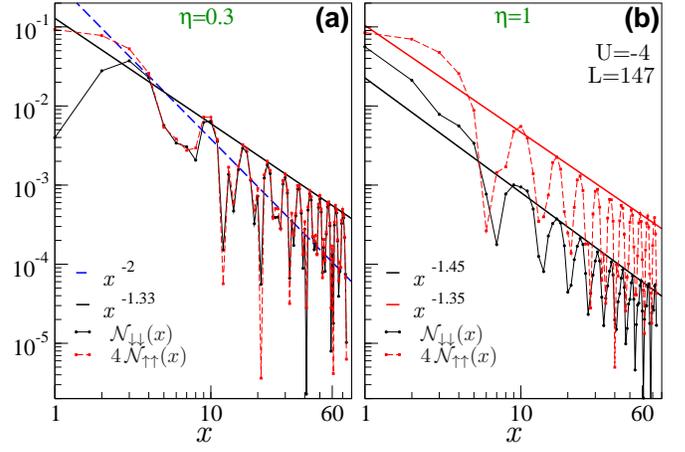}
\caption{(Color online). Density correlations in \textbf{(a)} the
  trimer phase and \textbf{(b)} the two-mode phase with $n=3/7$.}
\label{fig:densityCorr}
\end{figure}

\subsubsection{Locking between density correlations}

Lastly, the comparison of density-density correlations in the $\up$
and $\down$ channels is another interesting point of this model. In
fact, the bosonization results of Sec.~\ref{subsec:corr} predicts that
the exponent of $\mathcal{N}_{\up\up}(x)$ should be four times larger
than the exponent of $\mathcal{N}_{\down\down}(x)$ (if both remains
smaller than two), and that the dominant wave-vector should differ by
a factor two. Numerically, the typical behavior for a rather large
interaction $U=-4t_\up$ is given on Fig.~\ref{fig:densityCorr} for two 
values of the asymmetry $\eta$ in the single-mode and two-mode regimes. 
We see
that in the two-mode phase (Fig.~\ref{fig:densityCorr}\textbf{(b)}),
the two fluctuations have a slightly different exponent, that the
amplitude are quite different (including the natural factor four). Yet,
the dominant wave-vectors are both $2k_\up=k_\down$. In the trimer
phase, the disagreement with the bosonization picture is even worse
since the two densities are locked together, and nearly identical
(Fig.~\ref{fig:densityCorr}\textbf{(a)}). This latter fact cannot be
explained by the $1/x^2$ decay since the leading term is the
oscillating one, with an exponent clearly smaller than two. It is yet
physically not surprising in the strong coupling picture of
Fig.~\ref{fig:trimers} : trimers are local bound-states separated by
the typical distance $1/n_\up = 2\pi/2k_\up$ which does correspond to
the $2k_\up$ fluctuations but cannot be accounted by any of the
harmonics for the $\down$-density operator of Eq.~\eqref{HaldaneN} (we
work at an incommensurate filling). This short-distance binding cannot
be captured by the bosonization results of Sec.~\ref{sec:boso} but a
phenomenological Bose-Fermi approach described in the next section can
account for this strong-coupling regime. Lastly, the same comment can
be made on Friedel oscillations on the $\down$-component : they are
locked to the $\up$-component in the strong-coupling picture. One
might argue that there could be a crossover from the weak-coupling to
strong-coupling picture of Fig.~\ref{fig:trimers} as $|U|$ increases,
so that the bosonization results could be valid in the small $U$s
region. However, the trimer region is very sharp at small $U$s and we
could not find evidence for such a weak-coupling behavior in our
numerical data, although we cannot exclude such a possibility.

\subsection{Phenomenological Bose-Fermi picture at large $\vert{U}\vert$}
\label{subsec:bosefermi}

We here propose a simple picture that reconciles the numerical
observation and a bosonization approach at the prize of a strong
assumption, physically reasonable at large negative $U$, but difficult
to justify rigorously starting from the microscopic model.
This picture has been for instance proposed at large
interaction and low-density limit~\cite{Iskin2008}. 
Bose-Fermi mixtures of 1D models have extensively studied in recent
years~\cite{Cazalilla2003, 
Pollet2006, Mathey2007, Mathey2007a, Hebert2008, Barillier-Pertuisel2008, Rizzi2008, Mering2008,
Marchetti2009, Crepin2010, Orignac2010} and a similar picture emerges in certain
regimes of three-component Fermi-gases~\cite{Luscher2009}. When
$\vert{U}\vert$ is large, $\up$ and $\down$ fermions naturally form
onsite pairs which are effectively hard-core bosons which we
label $b$. We phenomenologically assume that the system is equivalent
to a Luttinger liquid of hard-core bosons with density $n_b = n_\up$,
an effective velocity $v_b$ and Luttinger parameter $K_b$, while the
remaining unpaired $\down$ fermions behave as a Luttinger liquid of
fermions labeled by $f$ and with parameters $n_f$, $v_f$ and
$K_f$. These two Luttinger liquids interact through an
effective interaction which will contain terms such as
\begin{equation}
\label{eq:BFcosine}
\int\! dx \, \cos{\left[2\pi(n_f-n_b)x - 2(\phi_f-\phi_b)\right]} \;,
\end{equation}
which have the tendency to lock the fields $\phi_f$ and $\phi_b$
together (with $\moy{\phi_f}=\moy{\phi_b}$ for attractive interaction), provided $n_f=n_b$. Such an effect has already been
discussed in the context of Bose-Fermi mixtures~\cite{Rizzi2008,
  Marchetti2009}. Clearly, the latter relation is the same as the
trimer commensurability condition $n_\down = 2n_\up$ (because bosons carry two particles) 
so that the
formation of trimer is now interpreted as a bound-state between the
bosons and the fermions. Following the same reasoning as in
Sec.~\ref{sec:fieldtransfo}, we can introduce a general transformation
of the $b/f$ fields into two new fields $c/d$ where $\phi_c =
(\phi_f-\phi_b)/\sqrt{2}$. Writing the matrix transformation from the
$r=c,d$ to the $s=b,f$ as $\mathfrak{p}_{rs}$ for the $\phi$s and
$\mathfrak{t}_{rs}$ for the dual $\theta$s, we have
\begin{align}
\label{eq:trs}
\mathfrak{t}_{cf} &= \frac{1}{\sqrt{2}}\;, & \mathfrak{t}_{cb} &= -\frac{1}{\sqrt{2}}\,.
\end{align}
which is slightly different from Eq.~\eqref{eq:tas}. Yet, the
2$k_F$-like fluctuating part of the density correlators for the
fermions and the bosons will have the leading contributions (dropping
the $x^{-2}$ terms) :
\begin{gather}
\mathcal{N}_{ff}(x) \sim \frac{\cos(2k_f x)}{x^{2\mathfrak{p}_{df}^2K_d}} \label{eq:fnupnup}\;, \\
\mathcal{N}_{bb}(x) \sim \frac{\cos(2k_b x)}{x^{2\mathfrak{p}_{db}^2K_d}} \label{eq:bfdndn}\;,
\end{gather}
which have both the same wave-vector associated with the Fermi levels
$k_f = k_b = k_\up$ and same decay exponents since $\mathfrak{p}_{df}
= \mathfrak{p}_{db}$ from the canonical transformation relations. In this picture, the
trimer is simply a bound-state between the bosons and the fermions so
its propagator reads
\begin{align}
T(x) &\sim e^{ik_f x}e^{i\theta_b}e^{i(\theta_f-\phi_f)} \\
&\sim e^{ik_\up x}e^{i[(\mathfrak{t}_{df}+\mathfrak{t}_{db})\theta_d 
+ (\mathfrak{t}_{cf}+\mathfrak{t}_{cb})\theta_c 
- \mathfrak{p}_{df}\phi_d - \mathfrak{p}_{cf}\phi_c]}\,.
\end{align}
From Eq.~\eqref{eq:trs} and the determinant of the transformation
matrix, which gives that $\mathfrak{t}_{df}+\mathfrak{t}_{db} =
1/\mathfrak{p}_{db} = 1/\mathfrak{p}_{df}$, we obtain that the
propagator is of the spinless fermionic type with an effective
Luttinger parameter $K'_{\text{eff}} = \mathfrak{p}_{df}^2K_d$. Clearly, both
the $f$-fermions and $b$-bosons propagators become short-range, the
latter corresponding to the pairing correlations in the native
fermionic model. Consequently, we recover the physics of the trimer
phase developed in Sec.~\ref{sec:boso}, with a better agreement with
the numerical observations in the strong coupling regime. However,
splitting the initial gas of $\down$ fermions into two parts can only
be done phenomenologically and could be questionable in a microscopic
derivation. This highlights the limitation of the bosonization
approach of Sec.~\ref{sec:boso} at short distances (high energies).

\subsection{Possible observation of the trimer phase in the presence of parabolic confinement}
\label{subsec:trapped}

\begin{figure}[b]
\centering
\includegraphics[width=\columnwidth,clip]{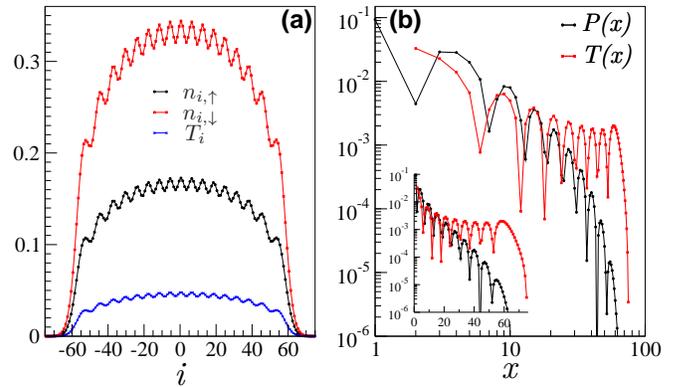}
\caption{(Color online). Realization of the trimer phase in a trapped
  configuration with parameters $U=-4t_\up$, $\eta=0.3$,
  $\omega=0.002\sqrt{t_\up}$, and $N=51$ fermions. \textbf{(a)} Local densities
  profiles. \textbf{(b)} Correlation functions from the center of the
  trap. \emph{Inset}: same in log-linear scale.}
\label{fig:trapped}
\end{figure}

In this section, we briefly discuss the condition to favor the trimer
phase in the presence of a parabolic confinement, as used in
cold-atoms experiments. Our goal is only to exhibit some parameters
for which the trimer phase is stabilized and to give some
qualitative comments. The trapping potential is taken into account
by adding the quadratic term
\begin{equation}
\mathcal{H}_{\text{trap}} = \frac 1 2 \omega^2 \sum_i (i-i_0)^2
\end{equation}
to Eq.~\ref{eq:hubbard}, with the trapping frequency $\omega$ and the
center of the lattice $i_0 = (L+1)/2$. According to a local-density
approximation (LDA) picture and using the phase diagram of
Fig.~\ref{fig:PhaseDiagU}, the trimer phase is likely to be found at
small enough densities and not too small $\eta$ to prevent the
occurrence of the FK regime. However, we find that the average density
of the trapped system is strongly dependent on the Hamiltonian
parameters. At fixed number of particles $N$ and trap size $\omega$,
changing $U$ and $\eta$ strongly affects the radius of the cloud and
the density in the middle. We only exhibit in Fig.~\ref{fig:trapped}
parameters for which the main features of the trimer phase are
reproduced in the presence of a parabolic confinement. The density
profiles of Fig.~\ref{fig:trapped}\textbf{(a)} illustrate the locking
of the $\up$ and $\down$ densities (up to exactly a factor two), and
the emergence of an appreciable density of local trimers (each local
maximum roughly corresponding to a trimer). In
Fig.~\ref{fig:trapped}\textbf{(b)}, the pairing and trimer
correlations are strongly different from that of a superfluid phase :
we have dominating trimer correlations and exponential pairing
correlations as in the homogeneous counterpart. In agreement with a LDA picture, since we have seen
that $K_{\text{eff}}$ decreases with density, the trimer correlations are boosted
at long distances. Similarly, the pairing correlations decrease
slightly faster then an exponential close to the edge of the
cloud. These results are encouraging in the perspective of a possible
achievement of the trimer phase in actual experiments.

\subsection{Observation of a crystal of trimers}
\label{sec:crystal}

\begin{figure}[t]
\centering
\includegraphics[width=\columnwidth,clip]{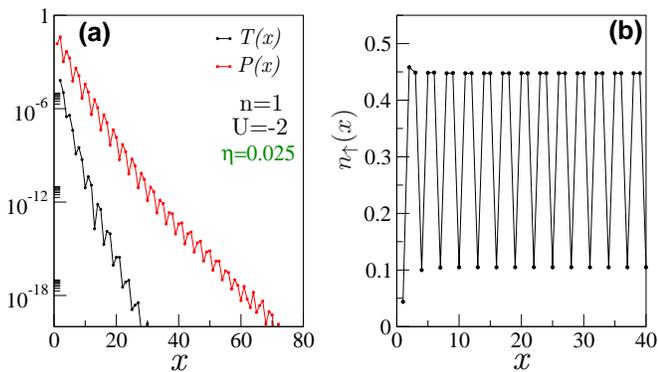}
\caption{(Color online). Observation of the crystallization of trimers
  in the commensurability situation of
  Fig.~\ref{latt:fig_tommaso} ($n=1$). \textbf{(a)} the short-range behavior of
  both pairing and trimer correlations. \textbf{(b)} ordering of the
  local density $n_\up$ with the expected period of three sites.}
\label{fig:crystal}
\end{figure}

According to the analysis of Sec.~\ref{sec:lattice}, a crystalline
phase of trimers can occur in this lattice model when the total
density $n$ is commensurate. Evidence of this scenario together with a
phase diagram for $n=1$ has been proposed in
Ref.~\onlinecite{Keilmann2009} in the case of a mixture of
two-component bosons, for large enough asymmetry. As the order
parameter (the density) associated with this transition is independent
of the statistics, we expect a similar scenario (see
Sec.~\ref{sec:lattice}) and a similar location of the transition in
the fermionic version of the model under study. Indeed, we give in
Fig.~\ref{fig:crystal} an example of the crystal phase. Notice that
very small $\eta$ are required to stabilize such a phase. We have not
investigated the extension of the phase which should rather be small
on the scale of the phase diagram of Fig.~\ref{fig:PhaseDiagU} and its
neighboring phases which could be either the two-mode LL or the
trimer phase. A crude argument can be proposed to understand this
crystallization within the Bose-Fermi picture of
Sec.~\ref{subsec:bosefermi} : when the mass asymmetry is very large
(very small $\eta$), it is reasonable to say that the mass of the
boson will be essentially the one of the heaviest particle, which is
the same as the unpaired fermions so that $v_f \simeq v_b$. In terms
of commensurability effects, one has $n_f = n_b = n/3$ so that
standard umklapp terms at $2\pi(n_f+n_b) = 4\pi/3$ do not account for
the crystallization. One rather has to look for higher order terms
with commensurabilities such as $2n_f+n_b = n_f+2n_b = n = 1$, which
are typically associated with terms like
\begin{equation}
\label{eq:BFcosinePlus}
\int\! dx \, \cos{2(2\phi_f+\phi_b)}\;,
\end{equation}
in addition to the one of Eq.~\eqref{eq:BFcosine}. Such term can lock
the field $2\phi_f+\phi_b$ and make the system fully gapped. Lastly,
we would like to stress that such commensurabilities are rather
surprising in terms of the initial fermion densities as they belong to
\emph{odd} filling fractions $n_\up = 1/3$ and $n_\down=2/3$.

\section{Conclusions}

In summary, the consideration of unusual commensurability conditions
in density-density interactions for 1D two-component
gases leads to a very rich physics with the possibility of building
bound-states of $(p+q)$-particles as the leading order. Such a
mode-locking mechanism can be described within the framework of
Luttinger liquid theory which reveals the main ingredients to
stabilize such a new phase. In particular, mass or velocity asymmetry
is shown to drive efficiently the transition into the multimer phase. 
Novel fully gapped phases are proposed
when taking into account umklapp couplings specific to lattice models
at commensurate densities. These ideas are illustrated and confronted with the
asymmetric 1D attractive Hubbard model for the special commensurability $2n_\up =
n_\down$ for which the formation of trimers is found. The features of
the phase diagram are computed, displaying the important role of the
density in favoring the trimer phase. The effective behavior of
trimers, which are effectively spinless fermionic objects, is very
sensitive to the density and mass asymmetry. Although the model seems
to have promising features to sustain a superfluid phase of trimers,
we did not find clear evidence for it in the thermodynamical limit,
while finite-size systems display a ``superfluid droplets'' physics.
Notice that superfluidity of bound-states made of four fermions
(quartets) can be achieved reliably in 1D with a four-color Hubbard
model~\cite{Capponi2008, Roux2009}. There, the bound-states are bosons
which natural ``free'' regime (attained in the low-density limit) is a
superfluid phase. A superfluid phase of trimers would in this respect
be even more exotic but is in strong competition with phase
separation. Lastly, we found that a trapping confinement supports the
trimer phase for reasonably high densities and that surprising crystal
phases can emerge at commensurate densities.

\begin{acknowledgments}
We would like to thank Giuliano Orso for earlier collaborations on related
problems. G.~R. thanks  Fran{\c c}ois Cr{\'e}pin, Fabian Heidrich-Meisner 
and Alexei Kolezhuk for fruitful discussions.
E.~B. gratefully acknowledges the hospitality of LPTMS, where the
majority of this work was done. We have benefited from the supports of the \textit{Institut
Francilien de Recherche sur les Atomes Froids} (IFRAF) and ANR
under grant 08-BLAN-0165-01.
\end{acknowledgments}

\end{document}